\documentclass[11pt,a4paper]{article}
\pdfoutput=1
\usepackage{jheppub}
\usepackage{amsmath,amssymb,amsfonts,bbold}
\usepackage{graphicx,bm}
\usepackage{verbatim}
\newcommand{\mma}{\texttt}

\title{Overdamped modes in Schwarzschild-de Sitter and a Mathematica package for the numerical computation of quasinormal modes}
\author{Aron Jansen}
\emailAdd{A.P.Jansen@uu.nl}
\affiliation{Institute for Theoretical Physics and Center for Extreme Matter and Emergent Phenomena, Utrecht University,\\ Leuvenlaan 4, 3584 CE Utrecht, The Netherlands}
\date{\today}

\abstract{
We present a package for Mathematica that facilitates the numerical computation of the quasinormal mode (QNM) spectrum of a black hole/black brane\cite{package}.
Requiring as input only the QNM equation(s), the application of a single Mathematica function will compute the spectrum efficiently, by discretizing the equation(s) and solving the resulting generalized eigenvalue equation.
It is applicable to a large variety of black holes, independent of their asymptotics.
The package comes fully documented and with several tutorials.
Here we present a self-contained review of the method and consider several applications.
We illustrate the method in the simplest case of scalar QNMs of a Schwarzschild black brane in anti-de Sitter.
Then we go on to look at the scalar QNMs of the Schwarzschild black hole in de Sitter, in anti-de Sitter and in asymptotically flat spacetimes, finding a novel infinite set of purely imaginary modes in the first case.
We also derive the QNM equations for a generic Einstein-Maxwell-scalar background and use these to compute the QNMs of the asymptotically anti-de Sitter Reissner-Nordstr\"{o}m black brane,
 as a further illustration and check of the method.
 }

\begin{document}
\maketitle
\section{Introduction}
Small perturbations of a black hole generically take the form of damped oscillations called quasinormal modes (QNMs).
These perturbations occur only at a discrete set of frequencies, depending on the black hole itself.
The spectrum of QNM frequencies contains a wealth of information about the black hole.

They describe the late time evolution of any dynamical process that results in a black hole at equilibrium.
In particular in the black hole mergers recently observed by LIGO\cite{LIGO1, LIGO2, LIGO3} the last part of the gravitational wave signal is described by quasinormal modes.
With more data of black hole mergers expected in the future, matching the signal to the quasinormal modes will be a good test of General Relativity\cite{LIGOtests}.
In particular as noted in \cite{Konoplya:2016pmh} a more accurate determination of the mass and angular momentum of the black hole is necessary to rule out or constrain many alternative theories of gravity.

More precisely, quasinormal modes are linear perturbations $\delta\phi$ which behave in time as
\begin{equation}
\delta\phi(t) \sim e^{-i \omega t} \, ,
\end{equation}
for some complex frequency $\omega$, its (negative) imaginary part giving exponential damping and its real part giving an oscillation.
By causality we require the perturbation to be ingoing at the black hole horizon, and either outgoing (asymptotically flat or de Sitter spacetimes) or finite (anti-de Sitter) at spatial infinity.
This results in a discrete spectrum of frequencies, $\omega_n$.

When there is no black hole there is nothing for the perturbation to dissipate energy into, there will be only normal modes, undamped oscillations, as we will see in \ref{sec:AdS}.

A perturbation can also become unstable, when $Im(\omega) > 0$, showing exponential increase instead of decay.
This can be used to look for new solutions as the end point of such instabilities, see e.g. \cite{reviewDiasSantosWay}.
There are also cases known where the instability has no end point and the black hole continues to grow until it hits the anti-de Sitter boundary \cite{scalarinstability,explosionInstability1,explosionInstability2}.

Through the fluid-gravity correspondence\cite{fluidgravity} the quasinormal modes of black branes are related to hydrodynamics.
In particular, the quasinormal modes whose frequencies vanish as momentum is taken to zero encode in principle all hydrodynamic transport coefficients in their dispersion relation \cite{KovtunStarinets}.
Higher modes decay more quickly and contain more information than hydrodynamics.
Recently \cite{resurgence} found strong evidence indicating that these modes set a limit on the range of applicability of hydrodynamics.

Holography relates the physics of black holes to the physics of strongly coupled quantum field theories\cite{holography}.
To each  field on the gravity side corresponds a gauge-invariant operator on the QFT side.
The quasinormal frequencies of this field are equal to the poles in the retarded Green's function of the operator in the dual QFT\cite{KovtunStarinets, SonStarinets}.
By summing quasinormal modes one can compute one loop determinants\cite{oneloop}.
Out of equilibrium gauge theory plasmas are dual to out of equilibrium asymptotically anti-de Sitter black holes.
At late times these are again described by quasinormal modes, see \cite{linearevolution, linearevolution2} for an explicit comparison.

For two extensive reviews on quasinormal modes, see \cite{reviewBertiCardosoStarinets} and \cite{reviewKonoplyaZhidenko}.


Being so full of information, it would be desirable to be able to compute these for any black hole.
While there are some cases where it can be done analytically (a recent quite involved example is \cite{analyticExample}, and see \cite{NatarioSchiappa} for a review of analytic computations and approximations), of course the generic case can only be done numerically.

In this paper we present a Mathematica package which numerically computes quasinormal modes\cite{package}. 
It is applicable to a broad class of cases:
it does not matter what the asymptotics are, if the equations are coupled, if the frequency occurs to a higher power in the equation, or if the background is numerical (assuming this numerical background has been computed).

The method used was first used for general relativity in \cite{Dias:2009iu,Dias:2010eu,Monteiro:2009ke}. 
Essentially it discretizes the quasinormal mode equation(s) using spectral methods, and then directly solves the resulting generalized eigenvalue equation.
Some great reviews on numerics in gravity are \cite{reviewCheslerYaffe, reviewDiasSantosWay, reviewGN}.

We have attempted to make the package easy to use, efficient in its computation, fully documented and with code that is itself easy to read and, if necessary, debug.
Earlier versions were used successfully for the quasinormal mode computations in \cite{scalarinstability,lifshitzevolution,entropy,RNdSSCC}. 
We encourage anyone to use it, and possibly to contribute by making improvements or adding new features.

The paper is structured as follows. 
In section \ref{sec:method} we describe the method used in the package.
Then in section \ref{sec:package} we describe how to use it at a more practical level, illustrated with one of the simplest cases: the Schwarzschild-anti-de Sitter black brane, in 5 dimensions.
The next two sections contain more examples. In section \ref{sec:schwarzschild} we study the 4-dimensional Schwarzschild black hole for various asymptotics: again anti-de Sitter, flat space and de Sitter.
In this last case we find an infinite set of purely imaginary modes that has been overlooked in the literature.
In section \ref{sec:EMD} we derive the quasinormal mode equations for a generic Einstein-Maxwell scalar action (with a homogeneous and isotropic background).
We use this to compute a more involved example: the 5-dimensional asymptotically anti-de Sitter Reissner-Nordstr\"{o}m black brane. 
This serves to illustrate the method in the more complicated case of coupled equations while also giving a check on the numerics, since we also have some analytic control on these quasinormal modes.
In the appendices we give a short report on the performance of the package and tables with numerical values for some quasinormal modes.

\section{Method}\label{sec:method}
To find the quasinormal mode spectrum, we have to solve the linear ordinary differential equation (ODE) (or coupled system of ODE's) describing a linearized perturbation on top of a black hole/black brane.
At the horizon, causality requires us to choose only ingoing waves.
Similarly at spatial infinity, or the cosmological horizon in the case of de Sitter, we must require only outgoing waves.
In asymptotically anti-de Sitter spacetime we require the perturbation to be normalizable at the boundary.

These two boundary conditions can only be satisfied for a discrete set of frequencies $\omega \in \{ \omega_n | n = 1, 2, 3, ...\}$\footnote{We stick to staring at $n=1$, though starting at $n=0$ is also common in the literature}, so at the same time as solving the ODE we have to solve for these $\omega_n$.

In this section we will discuss the method that is used in the package to do this.

\subsection{Analytics}\label{sec:analytics}
We will illustrate the method by the example of a massless scalar field in a 5-dimensional asymptotically anti-de Sitter Schwarzschild black brane.
It turns out that Eddington-Finkelstein coordinates, are perfectly suited for this problem.
For the time-independent backgrounds we will consider, these coordinates take the form
\begin{equation}\label{eq:EF}
ds^2 = - f(u) dt^2 + 2 g(u) dt du + \text{(spatial part)} \, .
\end{equation}
Although not strictly necessary, they usually simplify the problem, we will see why below, and we will use these coordinates throughout this paper. 

In these coordinates the asymptotically AdS Schwarzschild black brane is
\begin{equation}\begin{split}
f(u) &= u^{-2}(1-u^4) \, , \\
g(u) &= -u^{-2} \, ,
\end{split}\end{equation}
where $u=0$ corresponds to the boundary and $u=1$ corresponds to the horizon.

As long as the background is homogeneous, which in our case it is, we can write the fluctuation as a plane wave,
\begin{equation}\label{eq:qnm}
\delta \phi(u,t,x) = e^{-i \left(\omega t - k x \right)} \delta \phi(u) \, ,
\end{equation}
where $\omega$ is the frequency and $k$ is the momentum.
The same relation would hold for any other fluctuation in this background.

Inserting this into the Klein-Gordon equation we obtain,
\begin{equation}\label{eq:qnmads}
\left( -4 q^2 u -6 i \lambda \right) \delta\phi(u)+\left(-u^4+4 i \lambda u-3\right) \delta\phi'(u) + \left(u-u^5\right) \delta\phi''(u)= 0 \, .
\end{equation}
We have rescaled the quasinormal mode frequency $\omega$ and the momentum $k$ to the dimensionless ratios $\lambda \equiv \omega / 2 \pi T$ and $q \equiv k / 2 \pi T$, where $T = 1/(4\pi)$ is the temperature of the black brane.

Here we can already see one advantage of using Eddington-Finkelstein coordinates: the ansatz (\ref{eq:EF}) implies that $g^{tt} = 0$, and as a consequence the quasinormal mode equation \ref{eq:qnmads} is linear in the frequency, whereas it would be quadratic in Fefferman-Graham coordinates. The advantage of this will become clear in subsection \ref{sec:gev}.

Now we must analyze the behavior near the horizon and near the boundary to see how to deal with the boundary conditions.
Starting with the horizon, by plugging in an ansatz $\phi(u) = (1-u)^p$, we find that there are two solutions: $\delta\phi_{\text{in}}(u) = \text{const} + \mathcal{O}(1-u)$ and $\delta\phi_{\text{out}}(u) = (1-u)^{i \lambda} (1 + \mathcal{O}(1-u))$.

Including time dependence, this last solution behaves as 
\begin{equation}
\delta\phi_{\text{out}}(t,u) = e^{-i \lambda (2 \pi T t - \log(1-u) )} \, ,
\end{equation}
so as $t$ increases, $1-u$ has to increase as well to keep a constant phase, meaning $u$ has to decrease, which means that this solution is outgoing.
Hence we must make sure that we only get the other, ingoing solution.
Notice that the ingoing solution is perfectly smooth near the horizon, while the outgoing one oscillates more and more rapidly as we approach the horizon.
These properties will help us pick the correct solution, as we will see later.

Near the boundary there are two solutions, a non-normalizable mode $\delta\phi(u) \propto 1$ which we must discard, and a normalizable one $\delta\phi(u) \propto u^4$.
If we redefine $\delta\phi(u) \equiv u^3 \delta\tilde{\phi}(u)$, then the normalizable solution has $\tilde{\phi}$ approaching zero linearly, while the non-normalizable solution diverges.
\footnote{One can also rescale by $u^4$, but $u^3$ seems to be numerically more stable, the important point in any case is that the non-normalizable solution diverges while the normalizable one does not.}

Doing this rescaling, and also rescaling the equation itself so that it is finite but nontrivial at the boundary, the equation becomes:
\begin{equation}\label{eq:qnmfinite}
\left(-3-9 u^4 - 4 q^2 u^2 +6 i u \lambda \right) \delta\tilde{\phi}(u) +u \left(3-7 u^4+4 i u \lambda \right) \delta\tilde{\phi}'(u)+ \left(u^2-u^6\right) \delta\tilde{\phi}''(u)=0 \, .
\end{equation}
Now the normalizable solution behaves perfectly smoothly both at the boundary and at the horizon, while the non-normalizable solution behaves pathologically, diverging and rapidly oscillating respectively.
We will see in the next section how this can be used to automatically deal with the boundary conditions.

\subsection{Discretization: Pseudospectral Methods}\label{sec:spectral}
Having derived the equation (\ref{eq:qnmfinite}) in a form without divergences, we will now discretize it in order to solve it numerically.
For this we use pseudospectral methods, the standard reference on this is \cite{boyd}.
The pseudospectral method solves a (differential) equation by replacing a continuous variable, the radial one in our application, by a discrete set of points, also called collocation points. 
The collection of these points is usually called the grid.

A function can then be represented as the values the function takes when evaluated on the gridpoints. 
An equivalent and useful way of looking at this set of numbers representing a particular function is as coefficients of the so-called cardinal functions.
The cardinal functions corresponding to the grid $\{x_i | i = 0, ... , N \}$ are polynomials $C_j(x)$ of degree $N$, with $j = 0, ..., N$, satisfying $C_j(x_i) = \delta_{ij}$.
The choice of a grid uniquely specifies the cardinal functions as 
\begin{equation}
C_j(x) = \prod_{j=0,j\neq i}^{N} \frac{x - x_j}{x_i - x_j} \, .
\end{equation}

A function $f$ is then approximated as
\begin{equation}\label{eq:fapprox}
f(x) \approx \sum_{j=0}^N f(x_j) C_j(x) \, .
\end{equation}

The expansion in terms of cardinal functions allows one to construct the matrix $D_{ij}^{(1)}$ that represents the first derivative, $D_{ij}^{(1)} = C_i^\prime(x_j)$, and similarly for higher derivatives.

Solving the resulting linear equation will give a function which solves the original equation exactly at the collocation points. The hope is that as the number of gridpoints is increased, it will also solve the equation at other points.

For this to work, the choice of collocation points is crucial. The choice which tends to work best and is therefore most often used is the Chebyshev grid:
\begin{equation}\label{eq:grid}
x_i = \cos\left(\frac{i}{N} \pi \right) \, , i = 0, ..., N \, .
\end{equation}

For this grid it can be proven that any analytic function can be approximated with exponential convergence in $N$. 
This can be understood by realizing that when we increase $N$, not only does the largest distance between gridpoints decrease proportional to $1/N$, 
but at the same time the order of the cardinal functions increases, leading to a numerical error scaling as $(1/N)^N$.
Note that while this means that we can approximate the equation with exponential convergence, 
this does not mean that the number of quasinormal modes we find will grow exponentially with $N$, for more details see Appendix \ref{app:benchmarks}.

As given, these points lie in the interval [-1,1], but this can be rescaled and shifted to any other interval. In particular, we will want to rescale it to [0,1] if the horizon is at 1.

\begin{figure}[t]
\begin{center}
\includegraphics[width= \textwidth]{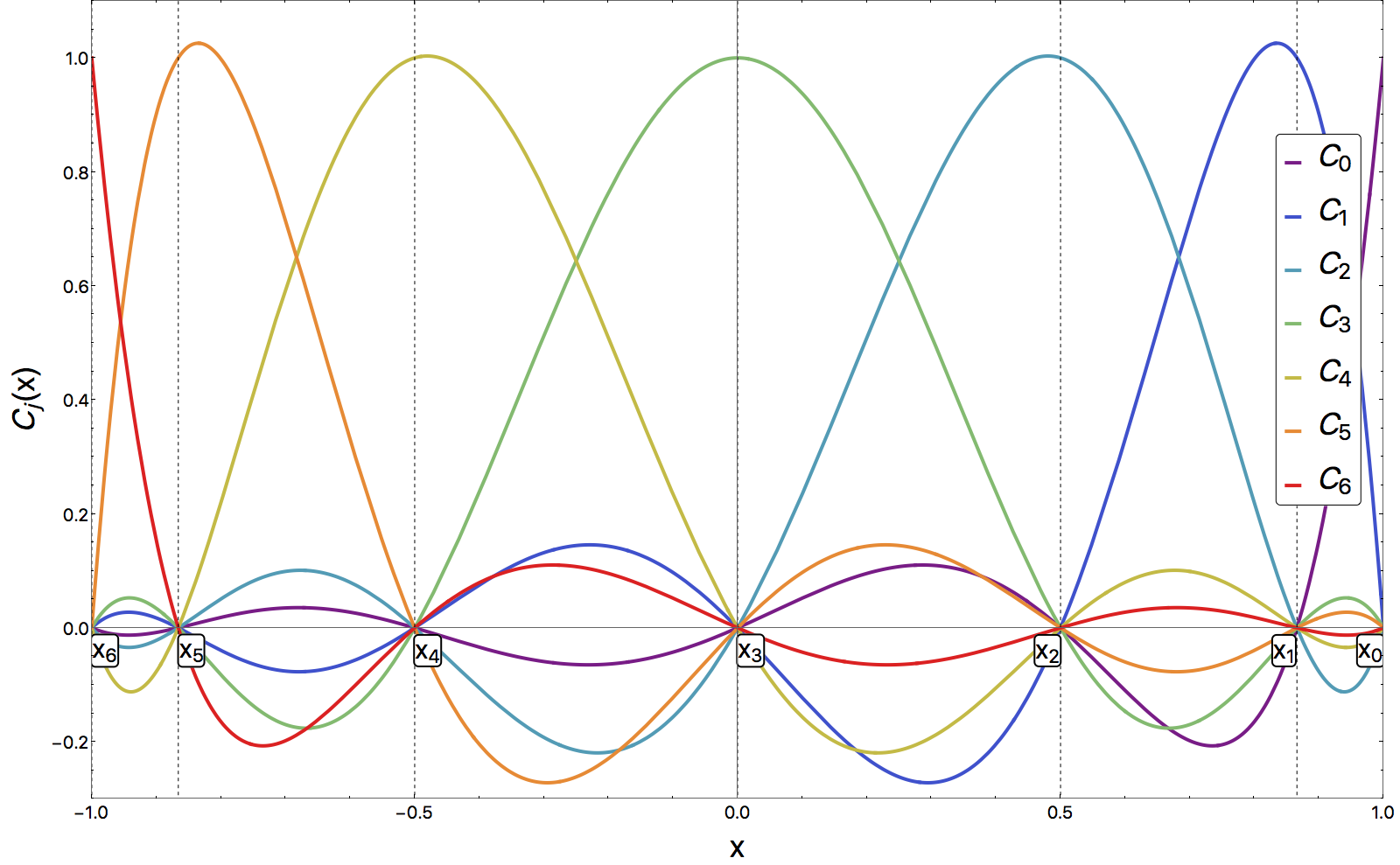}
\end{center}
\caption{Cardinal functions for the Chebyshev grid with $N=6$. Dashed lines indicate the grid points, at which by definition one of the cardinal functions equals 1, and all others vanish.}
\label{fig:cardinal}
\end{figure}

For this grid, the cardinal functions are linear combinations of Chebyshev polynomials $T_n(x)$:
\begin{equation}
C_j(x) = \frac{2}{N p_j} \sum_{m=0}^N \frac{1}{p_m} T_m(x_j) T_m(x) \, , \, \, \,  p_0 = p_N = 2 \, , \, \, p_j = 1 \, .
\end{equation}
In Fig. (\ref{fig:cardinal}) we show what these functions look like for $N = 6$.

Of course these functions are all perfectly smooth, and at the endpoints they are either $0$ or $1$ (since the endpoints themselves are collocation points).
With a linear combination of these smooth and finite functions we can never approximate either a function that is diverging or that is rapidly oscillating.
As a consequence, we have already implicitly solved the boundary conditions by choosing these functions as a basis.

\subsection{Generalized Eigenvalue Equation}\label{sec:gev}
Now we have turned the problem of solving a linear ODE, subject to specific boundary conditions, into solving a matrix equation, where the boundary conditions are already implicitly solved.
It is still not purely numerical though, as the equation depends on the frequency, which we have to solve for as well. 

This can be done by recognizing that this is a generalized eigenvalue problem.
The simplest type of quasinormal mode equation is of the form
\begin{equation}
c_0(u,\omega) \phi(u) + c_1(u,\omega) \phi^\prime(u) + c_2 (u,\omega) \phi^{\prime\prime}(u) = 0 \, ,
\end{equation}
where each of the $c_i$ are linear in $\omega$: $c_i(u,\omega) = c_{i,0}(u) + \omega c_{i,1}(u)$.

To be completely explicit, in our example Eq. (\ref{eq:qnmfinite}), setting the momentum $q$ to zero, this will give $c_{0,0} = -3 - 9 u^4$, $c_{0,1} = 6 i u$, $c_{1,0} = u(3 - 7 u^4)$, $c_{1,1} = 4 i u^2$, $c_{2,0} = u(u - u^5)$ and $c_{2,1} = 0$.
Each of these coefficients $c_{i,j}(u)$ is turned into a vector by evaluating it on the gridpoints.
These vectors are multiplied with the corresponding derivative matrices $D_{ij}^{(n)}$ and the resulting matrices added, to bring the equation into the form,
\begin{equation}\label{eq:gev}
\left(M_0 + \omega M_1 \right) \phi = 0 \, ,
\end{equation}
where the $M_i$ are now purely numerical matrices. Explicitly, $\left(M_0\right)_{ij} = c_{0,0}(x_i) \delta_{i j} + c_{1,0}(x_i) D_{ij}^{(1)} + c_{2,0}(x_i) D_{ij}^{(2)}$, and similarly for $M_1$.

Equation (\ref{eq:gev}) is precisely a generalized eigenvalue equation. This can be solved directly using Mathematica's built-in function Eigenvalues (or Eigensystem to get the eigenfunctions as well).

\subsection{Extensions}
The method presented works not just for the simple case used to illustrate it, in fact it works quite generally with a few simple modifications.

Firstly, suppose that instead of one ODE, we have a coupled system of ODE's. Say there are two of them (but the following applies generally), of the form
\begin{equation}\begin{split}
c_0(u,\omega) \phi(u) + c_1(u,\omega) \phi^\prime(u) + c_2(u,\omega) \phi^{\prime\prime}(u) + d_0(u,\omega) \psi(u) + d_1(u,\omega) \psi^\prime(u) + d_2(u,\omega) \psi^{\prime\prime}(u) &= 0 \, , \\
e_0(u,\omega) \psi(u) + e_1(u,\omega) \psi^\prime(u) + e_2(u,\omega) \psi^{\prime\prime}(u) + f_0(u,\omega) \phi(u) + f_1(u,\omega) \phi^\prime(u) + f_2(u,\omega) \phi^{\prime\prime}(u) &= 0 \, .
\end{split}\end{equation}

We can discretize this in a similar way, by joining the two functions $\phi$ and $\psi$ into a single vector $(\phi(u),\psi(u))$.
The matrix $M_0$ of Eq. (\ref{eq:gev}) becomes
\begin{equation}
M_0 = \begin{pmatrix} \tilde{M}_{0,1}(\phi) & \tilde{M}_{0,1}(\psi) \\  \tilde{M}_{0,2}(\phi) & \tilde{M}_{0,2}(\psi) \end{pmatrix} \, ,
\end{equation}
where the $0$ everywhere indicates that this is the piece of order 0 in the frequency, the second index indicates the equation number and the argument indicates the function, so that $\tilde{M}_{0,1}(\phi)$ is the matrix coefficient of $\phi$ in the first equation.
Of course there is a similar equation for the linear term in the frequency.

Further, we can generalize to an equation depending on the frequency as an arbitrary polynomial.
Whether coupled or not, the procedure above will bring such a (system of) equation(s) to the form
\begin{equation}
\mathcal{M}(\omega) \phi = \left( \tilde{M}_0 + \omega \tilde{M}_1 + \omega^2 \tilde{M}_2 + ... + \omega^p \tilde{M}_p \right) \phi = 0 \, .
\end{equation}
Note that in the case of a coupled equation, the $\phi$ here would be a vector composed of several functions, as above.

We can again write this as a generalized eigenvalue equation of the form of Eq. (\ref{eq:gev}) by defining,
\begin{equation}
M_0 = \begin{pmatrix}
 \tilde{M}_0 & \tilde{M}_1 &  \tilde{M}_2 & \cdots & \tilde{M}_{p-1} \\
 0 & \mathbb{1} & 0 & \cdots & 0 \\
 0 & 0 & \mathbb{1} & \cdots & 0 \\
 \vdots & \vdots & \vdots & \ddots & \vdots \\
 0 & 0 & 0 & 0 & \mathbb{1} \\
 \end{pmatrix} \, , \, \, \, 
 M_1 = \begin{pmatrix}
 0 & 0 & 0 & \cdots & 0 & \tilde{M}_{p} \\
 -\mathbb{1} & 0 & 0 & \cdots & 0 & 0 \\
 0 & -\mathbb{1} & 0 & \cdots & 0 & 0 \\
 \vdots & \vdots & \vdots & \ddots & \vdots & \vdots \\
 0 & 0 & 0 & \cdots & -\mathbb{1} & 0 \\
 \end{pmatrix} \, .
\end{equation}
These matrices act on the vector $(\phi^{(0)},\phi^{(1)},\phi^{(2)}, ..., \phi^{(p-1)})$, where the first row represents the original equation, and the other rows enforce that $\phi^{(i)} = \omega^i \phi$.

Notice that for a coupled system of $N_{\text{eq}}$ equations with the maximal power of the frequency being $p$, the resulting matrix will be $N_{\text{eq}} p (N+1)$ (where $N$ is the $N$ in Eq. (\ref{eq:grid})).
So both extensions come at a price in computational time.

An alternative here is to solve $\text{det}(\mathcal{M}(\omega)) = 0$. This is prohibitively slow to do symbolically, but it can be done numerically by sweeping the complex $\omega$-plane.
This has the advantage that higher powers of $\omega$ don't require larger matrices, but the extra complexity of selecting and possibly refining a grid of points in the complex plane.
\footnote{This is also implemented in the package (by setting \mma{Method$\rightarrow$``Sweep''}), but not as worked out as the main method.}

\section{Using the package}\label{sec:package}
In this section, we will show at a very practical level how to work with the package.
Our starting point will be the properly rescaled equation for a massless scalar in an asymptotically anti-de Sitter Schwarzschild black brane, at zero momentum, Eq. (\ref{eq:qnmfinite}).

First some practical remarks. The package can be found here \cite{package}, where installation instructions can also be found.
Once installed there are several ways to get started and familiarize oneself with it. 
Apart from the present paper, the package also has its own documentation. 
This can be accessed through Mathematica by going to Help, Documentation and typing in ``QNMspectral'', and describes all the functions and their options.
It includes some tutorials, which also go into more detail on how to obtain Eq. (\ref{eq:qnmfinite}).
We also provide a separate notebook with several examples, also found at \cite{package}.

\begin{figure}[h]
\begin{center}
\includegraphics[width=.7\textwidth]{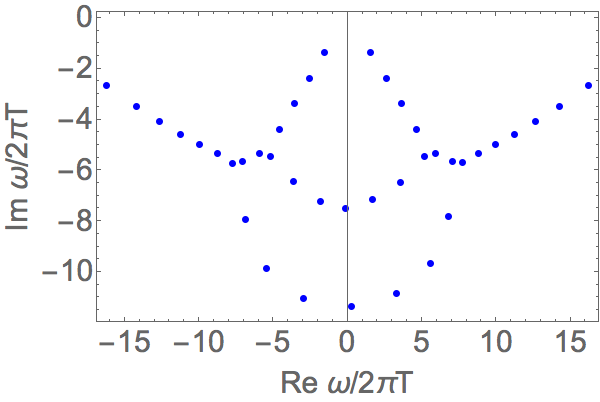}
\end{center}
\caption{Eigenvalues found from evaluating (\ref{eq:getmodes}). Using a grid with 41 points, 41 eigenvalues are found, only a fraction of which actually correspond to quasinormal modes, the rest are numerical artifacts.}
\label{fig:modes1}
\end{figure}

Here we continue with Eq. (\ref{eq:qnmfinite}).
The function which implements everything mentioned in the previous section is called \mma{GetModes}.
Assuming Eq. (\ref{eq:qnmfinite}) is stored in Mathematica under \mma{eq}, one can compute the quasinormal modes simply as
\begin{equation}\label{eq:getmodes}
\mma{\text{modes = GetModes[eq,\{40,0\}];} }
\end{equation}
This does the computation with $N=40$ (meaning with 41 gridpoints), at machine precision (machine precision has about 16 digits, anything less than that in the last argument will use machine precision).
Note that it is not necessary to specify what are the frequency, the radial variable and the fluctuation, this can be determined unambiguously from the equation and is done automatically.

Now the result, a list of quasinormal mode frequencies, is stored in \mma{modes} and can be displayed by evaluating \mma{PlotFrequencies[modes,Name$\rightarrow$``$\omega / 2 \pi T$'']}, producing Fig. (\ref{fig:modes1}).

The computation used a grid of 41 points, so there are 41 eigenvalues.
Typically the quasinormal modes lie in an approximately straight line, they certainly should in this case. 
This means that most of the eigenvalues found are numerical artifacts, only a few are accurate.
This is unavoidable and will continue to be the case if we increase $N$ or the precision.
We will get more accurate results, but also more results in general, so still only a small fraction of the total will be accurate.

To test whether a computed eigenvalue really is a quasinormal mode and not just a numerical artifact, one has to repeat the computation at different grid sizes and precisions, and look for convergence.
A simple and efficient implementation of this is given in the function \mma{GetAccurateModes}. Used for example as
\begin{equation}\label{eq:getaccmodes}
\mma{\text{modesaccurate = GetAccurateModes[eq,\{40,0\},\{80,40\}];} }\,  ,
\end{equation}
this does the computation twice, once with $N = 40$ at machine precision and once with $N = 80$ at precision 40. It returns only those modes which occur in both computations, and of the modes it returns, it takes only those digits which agree.

\begin{figure}[htb]
  \centering
  \begin{minipage}[c]{0.5\textwidth}
    \centering
    \includegraphics[width=\textwidth]{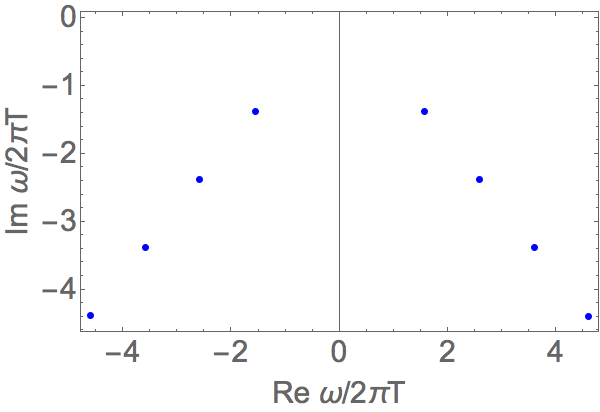}
  \end{minipage}
  \begin{minipage}[c]{0.4\textwidth}
    \centering
\begin{tabular}{c|c|c|}
n & Re $\omega_n/2 \pi T$ & Im $\omega_n / 2 \pi T$ \\ \hline
1 & $\pm 1.559725796$  & -1.373337872 \\ \hline
2 & $\pm 2.584760$ & -2.381785 \\ \hline
3 & $\pm 3.594$ & -3.385 \\ \hline
4 & $\pm 4.6$ & -4.4 \\ \hline
\end{tabular}
  \end{minipage}
  \caption{Eigenvalues found from evaluating (\ref{eq:getaccmodes}). These are the modes that were correctly computed using a grid with $N=40$ and machine precision, as judged by a comparison with a more accurate computation with a grid with $N = 80$ using 40 digits of precision. On the right is a table showing the exact numerical values. Assuming this comparison is a good criterion for convergence, all digits shown are accurate.}
\label{fig:modesacc}
\end{figure}

Note that for this to actually give only converged modes, one has to be sure that the equation itself is either fully analytic, or if it is numeric that the error in the equation is not the dominant one. 
One also has to be sure that there is no common error in both computations, as this will not get filtered out.
This can happen when taking the grid sizes too close.

We can display these results in a plot of the complex plane and a table by evaluating \mma{ShowModes[modesaccurate,Name$\rightarrow$``$\omega/2\pi T$'',Precision$\rightarrow$Infinity]}, giving Fig. (\ref{fig:modesacc}).
The option \mma{Precision$\rightarrow$Infinity} is used here to show all computed digits, by default only the first 6 are shown.\footnote{For more details on the various options of all the functions in the package see the documentation, which can also be accessed by evaluating for example \mma{?GetAccurateModes}, as with built-in Mathematica functions.}
These values can be compared with \cite{planarAdS}, where we see that indeed all displayed digits are accurate.

We see that the lowest mode has been computed quite accurately, and as the mode number increases the precision of our result decreases. 
Apart from the caveats mentioned above, in practice this often gives only correct results. We advise however to do more elaborate checks to ensure accuracy.

\begin{figure}[h]
\begin{center}
\includegraphics[width=.6\textwidth]{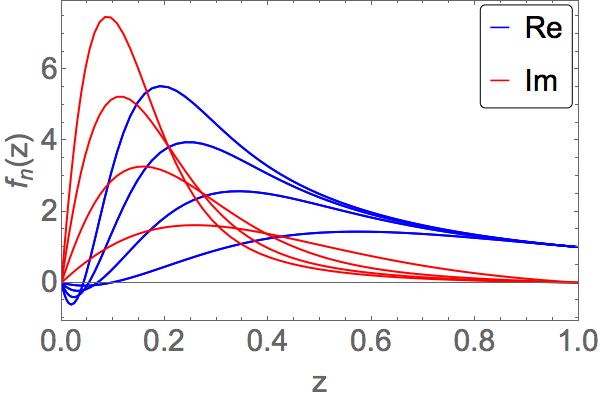}
\end{center}
\caption{Eigenfunctions obtained from evaluating \mma{GetAccurateModes[eq,$\{40,0\},\{80,40\}$,} \mma{Eigenfunctions$\rightarrow$True]}. Eigenfunctions come in conjugate pairs, only those with positive imaginary part are shown. All are smooth, normalized to go to 1 at the horizon, and go to 0 at the boundary as was expected from the rescaling that we did.}
\label{fig:eigenfuncts}
\end{figure}

One additional check that one can do is to look at the eigenfunctions as well, and check that they are smooth and finite.
By default, these are not computed. To compute them, simply add the option \mma{Eigenfunctions$\rightarrow$True} to either \mma{GetModes} or \mma{GetAccurateModes}.
Repeating the computation (\ref{eq:getaccmodes}) with this option, we can now plot the eigenfunctions by \mma{PlotEigenfunctions[modesaccurate]}, producing Fig. (\ref{fig:eigenfuncts}).

In this case, one can see that all eigenfunctions are perfectly smooth, and go to 0 at the boundary, as was expected from the rescaling we did. 
Notice that they are normalized to be 1 at the horizon.

\section{Schwarzschild black hole}\label{sec:schwarzschild}
As another simple example and to illustrate the general applicability we now consider the fluctuations of a massless scalar on top of a Schwarzschild black hole in various 4-dimensional spacetimes: anti-de Sitter, flat space and de Sitter.
These black holes are all maximally symmetric solutions of the equations of motion coming from the action
\begin{equation}\label{eq:action}
S = \frac{1}{8 \pi G}\int d^4 x \sqrt{-g} \left( R - 2 \Lambda \right) \, .
\end{equation}

The background metric we take is
\begin{equation}\begin{split}\label{eq:bgschw}
ds^2 &= - f(u) dt^2 + 2 \zeta(u) dt dr + S(u)^2 \left( d\theta^2 + \sin^2 (\theta) d\phi^2 \right) \, , \\
\zeta(u) &= - u^{-2} \, , \\
S(u) &= u^{-1} \, ,
f(u) = 1 - 2 G M u + \epsilon \frac{1}{L^2 u^2} \, , 
\end{split}\end{equation}
where we set Newton's constant $G = 1$, $M$ is the mass of the black hole, and depending on the asymptotics we have $\epsilon = 1, 0, -1$ and $\Lambda = -3/L^2, 0, 3/L^2$, for anti-de Sitter, flat space and de Sitter respectively.

Apart from the different asymptotics to the previous example we take a different horizon topology: a sphere instead of a plane.
This has the consequence that plane waves are no longer solutions, and we must instead use spherical harmonics, labelled by a multipole number $l$.
The perturbation we can do is then:
\begin{equation}\label{eq:qnmds}
\delta \phi(t,u,\theta,\phi) = e^{- i \omega t} Y_l{}^m(\theta,\phi) \delta\phi(u) \, .
\end{equation}

These perturbations have to satisfy the Klein-Gordon equation, which takes the form:
\begin{equation}\label{eq:KGd4}
\left(u l (l+1) +2 i \omega \right) \delta\phi (u) -  \left(2 i u \omega + u^3 f'(u) \right)  \delta\phi '(u) - u^3 f(u) \delta\phi ''(u) = 0 \, .
\end{equation}
In each of the cases considered, the behavior of the fluctuation near the horizon is the same as in the black brane discussed before.
There is an outgoing mode $\delta\phi_\text{out}(u) = (1-u)^{i \lambda}$, and an ingoing mode $\delta\phi_{\text{in}}(u) = \text{const.}$, due to our use of Eddington-Finkelstein coordinates.
This is perfect, as the numerical approximation is only able to resolve the ingoing mode.

We will explore each of the cases in turn, starting with the one most similar to the previous example: anti-de Sitter.

\subsection{Anti-de Sitter}\label{sec:AdS}
Just as in the previous example, anti-de Sitter has a conformal boundary at $u = 0$.
The scalar field contains a normalizable mode as well as a non-normalizable mode, behaving near the boundary respectively as $\phi \propto u^3$ and $\phi \propto \text{const}$.
We do not want our fluctuation to change the boundary, therefore we demand the non-normalizable mode to vanish.
Similar to before, we do this by redefining 
\begin{equation*}
\delta \phi(u) = u^2 \tilde{\delta\phi}(u) \, ,
\end{equation*}
so that the normalizable mode approaches 0 as $u \rightarrow 0$ and the non-normalizable mode diverges.

The blackening factor, in Eq. (\ref{eq:bgschw}) with $\epsilon = 1$, has a horizon at some $u = u_h$, in terms of which the mass is $M = (1 + L^2 u_h^2)/(2 L^2 u_h^3)$ and the black hole temperature is $T = (3 + L^2 u_h^2)/(4 \pi L^2 u_h)$.
We will set $u_h = 1$ without loss of generality. 
Replacing $M$ by $L$ using the equation above, replacing $\omega$ by $\lambda \equiv \omega / (2 \pi T)$ and doing the rescaling of $\delta\phi$ we obtain the quasinormal mode equation:
\begin{equation}\begin{split}\label{eq:globalads}
0 =&\left( -2 - 4 u^3 + (2-4 u - l (l+1)) u^2 L^2 + \left(L^2+3\right) u i \lambda \right) \tilde{\delta\phi}(u) + \\
& u \left( 2 - 5 u^3 +  (4 - 5 u) u^2 L^2 + \left(L^2+3\right) u i \lambda \right) \tilde{\delta\phi}^\prime(u) + \\
&(1-u) u^2 \left(1 + u + \left(L^2+1\right) u^2\right) \tilde{\delta\phi}^{\prime\prime}(u) \, .
\end{split}\end{equation}
Note that we now have the dimensionless ratio $M/L$ as a parameter describing the relative size of the black hole and the whole anti-de Sitter universe.

\begin{figure}[t]
  \centering
  \begin{minipage}[c]{0.49\textwidth}
    \centering
    \includegraphics[width=\textwidth]{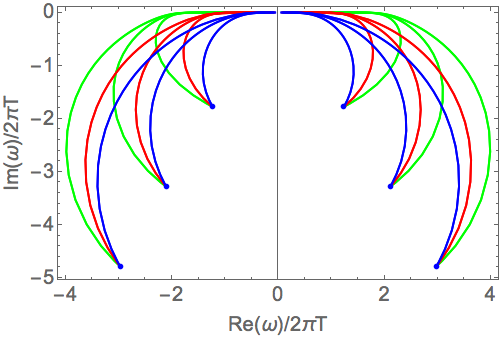}
  \end{minipage}
  \begin{minipage}[c]{0.49\textwidth}
    \centering
    \includegraphics[width=\textwidth]{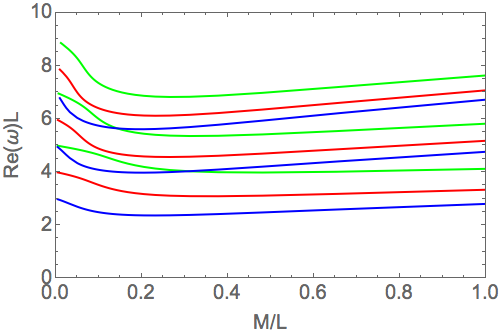}
  \end{minipage}  
\caption{Quasinormal modes of a massless scalar of a Schwarzschild black hole in global anti-de Sitter spacetime, for multipole numbers $l = 0$ (blue), $l=1$ (red) and $l=2$ (green).
 Left: plotted on the complex plane. Right: real part as a function of $M/L$, with the imaginary part approaching 0 as $M/L \rightarrow 0$.}\label{fig:SAdS}
\end{figure}

In Fig. (\ref{fig:SAdS}) we show the quasinormal modes for $l = 0, 1, 2$ as $M/L$ is varied between 0 and infinity.
Some of these values were computed before in \cite{HubenyHorowitz}, with which we of course agree\footnote{The case where the scalar also has a conformal coupling was studied in \cite{confcoupling1,confcoupling2}.}.

In the limit $M/L \rightarrow \infty$ the Schwarzschild black brane is approached.
The $l$-dependence drops out and all modes converge at those of the brane, indicated by the blue dots.
The opposite limit $M/L \rightarrow 0$ is that of empty anti-de Sitter. In this limit we obtain the normal modes of empty anti-de Sitter, $\omega_n L = \pm( l + 1 +2 n) $, $n = 1, 2, ...$ \cite{pureAdS, pureAdSlimit, HubenyHorowitz}.
These are pure oscillations without decay, since there is no horizon to decay into.

\subsection{Flat Space}\label{sec:flat}
Now we discuss the case of flat asymptotics.
The background is obtained by setting $\epsilon = 0$ in Eq. (\ref{eq:bgschw}), which when inserted into Eq. (\ref{eq:KGd4}) gives the quasinormal mode equation for the asymtotically flat Schwarzschild black hole:
\begin{equation}\label{eq:flatSchw0}\begin{split}
0 &=\left( u\, l (l+1)+2 i  \omega \right) \phi (u) +\left(u^3-2 u \, i  \omega \right) \phi '(u)-(1-u) u^3 \phi ''(u) \, .
\end{split}\end{equation}

The behavior of $\phi$ is singular at spatial infinity, $u = 0$, but we can scale this out.
There are again two solutions, $\phi_{\text{out}}(u) = \exp\left(2 i \omega/u\right) u^{1-2 i \omega}( 1 + ...)$ representing waves going out to infinity, and $\phi_{\text{in}}(u) = u (1 + ... )$ representing waves coming in from infinity.
We must demand the latter to vanish, so we redefine
\begin{equation}\label{eq:bcflat}
\delta\phi(u) = \exp \left(2 i \omega/u\right) u^{-2 i \omega} \tilde{\delta\phi}(u) \, ,
\end{equation}
so that $\tilde{\delta\phi}(u) \propto u$ for the outgoing modes, while the modes coming in from spatial infinity diverge.

We set the horizon $u_h = 1/(2M) = 1$ and define $\lambda = \omega M$, to get the final equation,
\begin{equation}\label{eq:flatSchw}\begin{split}
0 =& (l (l + 1) u - 4 i \lambda - 16 u (1 + u) \lambda^2) \tilde{\phi}(u) + (u^3 + 4  u (1 - 2 u^2) i \lambda) \tilde{\phi}^\prime(u) \\
&+ (-1 + u) u^3 \tilde{\phi}^{\prime\prime}(u) \, .
\end{split}\end{equation}

\begin{figure}[t]
\begin{center}
\includegraphics[width=\textwidth]{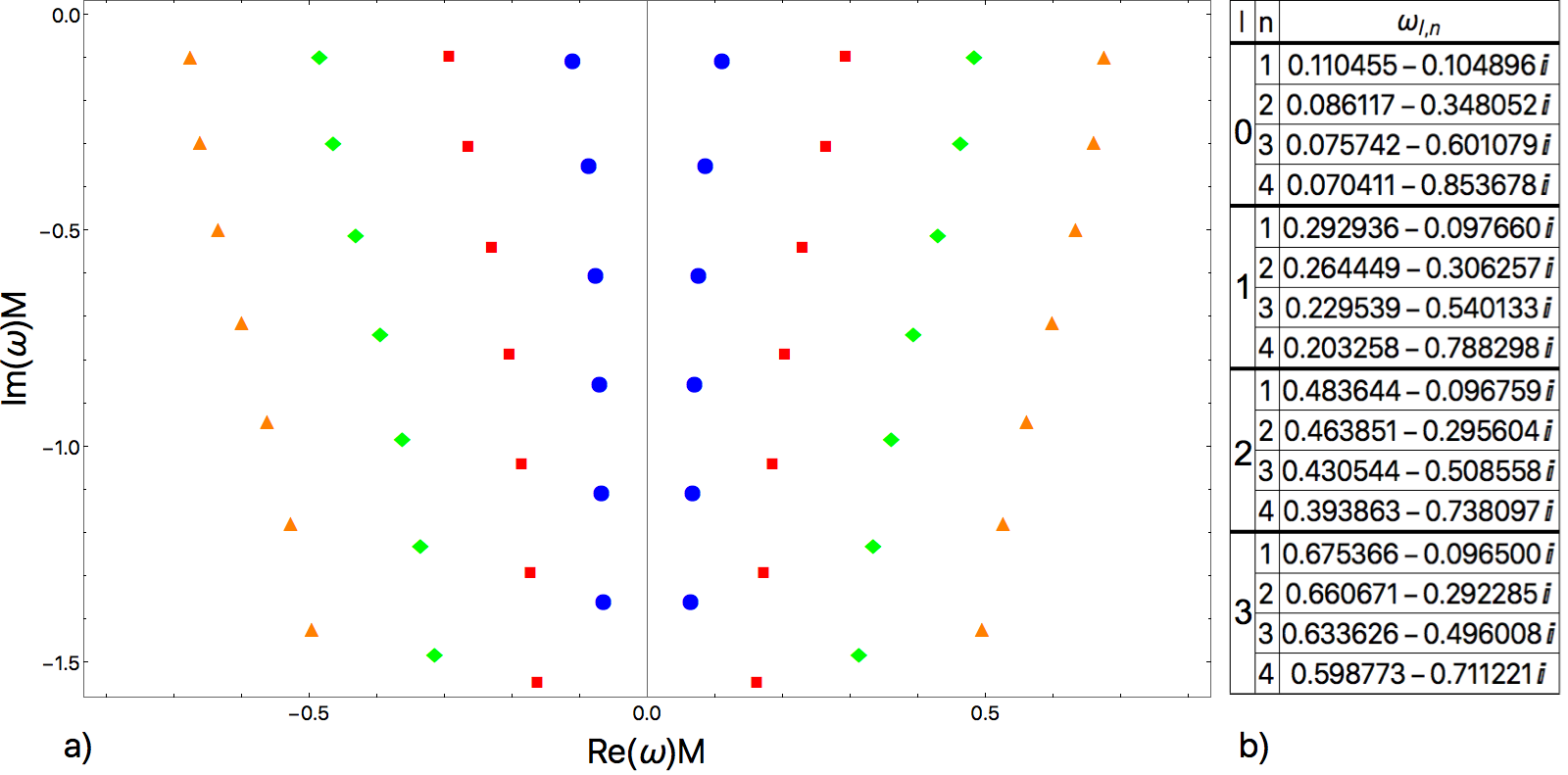}
\end{center}
\caption{
Scalar quasinormal modes of the Schwarzschild black hole in asymptotically flat spacetime, for multipole number $l = 0$ (blue circles), $l = 1$ (red squares), $l=2$ (green diamonds) and $l=3$ (orange triangles).
On the right is a table with the lowest 4 modes for each $l$.
}
\label{fig:flatmodes}
\end{figure}

In Fig. \ref{fig:flatmodes} we show the scalar quasinormal modes for $l = 0, 1, 2, 3$, along with a table for the first four modes for each $l$.
These points reproduce table 1 of  \cite{flatS} exactly. 

\subsection{De Sitter}\label{sec:SdS}
The Schwarzschild-de Sitter black hole is the most physically rich case of all.
There are two horizons: the cosmological de Sitter horizon and the black hole horizon, whose coordinates we will denote as $u_c$ and $u_b$ respectively.
The blackening factor, Eq. (\ref{eq:bgschw}) with $\epsilon = -1$, can be conveniently rewritten in terms of these quantities as 
\begin{equation}\label{eq:fSdS}
f(u) = 1 - \frac{u_c^2 u_b^2}{u_c^2+u_c u_b + u_b^2} \frac{1}{u^2} - \frac{u_c+u_b}{u_c^2+u_c u_b + u_b^2} u \, .
\end{equation}
Note that this is symmetric in $u_c$, $u_b$, but for our naming to make sense we require that $u_c < u < u_b$, so in particular we have to restrict $0 < u_c < u_b$.

These horizons have surface gravities $\kappa_c = | u_c^2/2 f^\prime(u_c) |= \frac{u_c}{2} \frac{ 2 u_b^2 - u_c u_b - u_c^2}{u_c^2+u_c u_b + u_b^2}$ and
$\kappa_b = | u_b^2/2 f^\prime(u_b) | =  \frac{u_b}{2} \frac{ u_b^2 + u_c u_b -2 u_c^2}{u_c^2+u_c u_b + u_b^2}$ respectively.
The quasinormal mode equation in these parameters becomes,
\begin{equation}\begin{split}\label{eq:dsqnm0}
 0= & \left(u_c^2+u_b u_c+u_b^2\right) \left(u l (l+1)+2 i \omega \right)  \phi (u)+\\
 & \left( u^3 \left(u_b+u_c\right)-2 i u \omega  \left(u_c^2+u_b u_c+u_b^2\right)-2 u_b^2 u_c^2 \right) \phi '(u) +  \\
 &u \left(u-u_b\right) \left(u-u_c\right) \left(u_b u_c+u \left(u_b+u_c\right)\right) \phi ''(u) \, .
\end{split}\end{equation}

The background solution has two independent dimensionful parameters, from which we can extract a dimensionless ratio which we will take as $M/L$, the mass of the black hole divided by the radius of de Sitter.
In terms of $u_c$ and $u_b$ this is 
\begin{equation}\label{eq:MoverL}
M/L = \frac{u_b u_c \left(u_b+u_c\right)}{2 \left(u_c^2+u_b u_c+u_b^2\right){}^{3/2}}  \, .
\end{equation}

Solving Eq. (\ref{eq:dsqnm0}) near the cosmological horizon we again find two solutions,  $\delta\phi(u) = (u-u_c)^{ - i \lambda_c}$, with $\lambda_c \equiv \omega / \kappa_c$, or $\delta\phi(u) \rightarrow \text{const}$.
Restoring time dependence on the first solution we get $\delta\phi(t,u) = \exp\left(- i \omega (t + \frac{1}{2\pi T_c} \log(u - u_c) ) \right)$. 
To keep a constant phase as $t$ increases, we need to decrease $u$, which means that this solution is going into the cosmological horizon.
In order to keep only this solution we need it to be smooth while the outgoing one oscillates rapidly, so we have to redefine
\begin{equation}\label{eq:dsrescale}
\delta\phi(u) = \frac{1}{u-u_c} \left(u-u_c \right)^{-i \lambda_c} \tilde{\delta\phi}(u) \, ,
\end{equation}
so that $\tilde{\phi}(u) \propto (u-u_c)$ for the ingoing solution, while the outgoing solution oscillates rapidly.

Making this replacement, rescaling the radial coordinate $u_c < u < u_b $ to $0 < x < 1$, and setting $u_b = 1$, we obtain the final quasinormal equation, shown in appendix \ref{sec:app2}.

\begin{figure}[t]
  \centering
  \includegraphics[width=\textwidth]{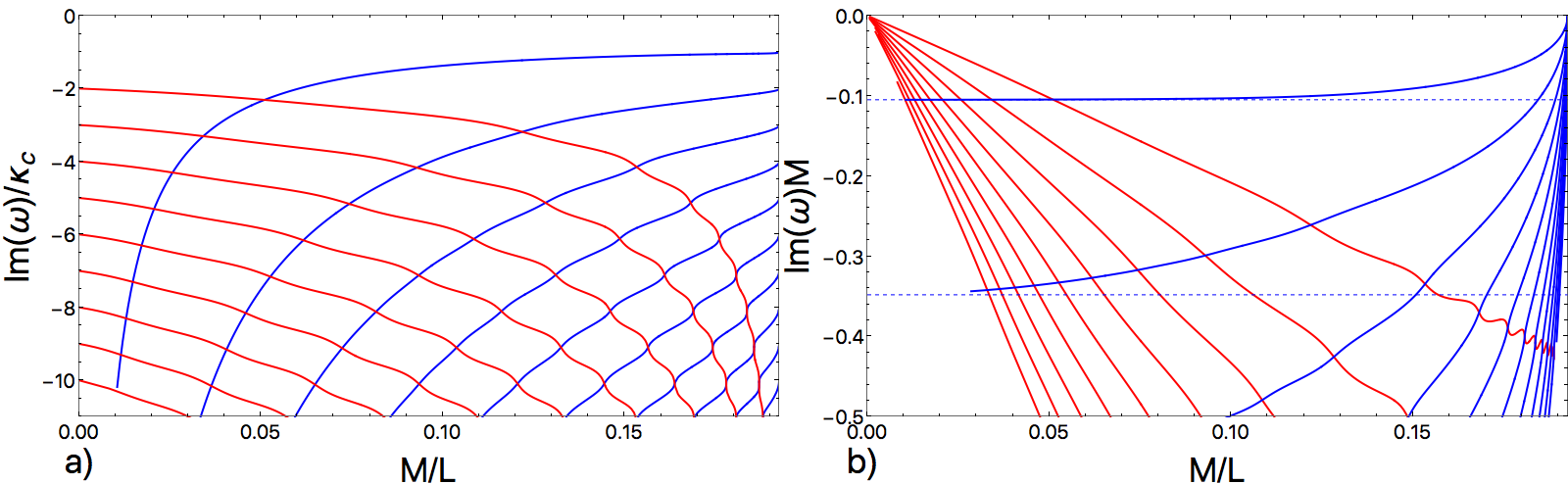}
  \caption{
  Massless scalar quasinormal modes of the asymptotically de Sitter Schwarzschild black hole in 4 dimensions for $l=0$, as a function of $M/L$.
  Red lines are purely imaginary modes, while blue lines have a real part as well. 
  Normalized as $\omega/\kappa_c$ on the left, showing clearly the limits of Eqs. (\ref{eq:extremalSdS},\ref{eq:dSanalytic}), and normalized as $\omega M$ on the right, showing the limit of the asymptotically flat Schwarzschild black hole, indicated with dashed lines.
  For clarity we show only the first nine overdamped modes (even though all should appear in the right plot).
  }
  \label{fig:SdSim}
\end{figure}

Before going into the numerical results it is instructive to look at the limiting cases. 
One limit is the extremal limit $M/L \rightarrow 1/(3 \sqrt{3})$. In this limit $u_c \rightarrow u_b = 1$, but the proper distance remains finite and one obtains the Nariai spacetime.
Here the quasinormal modes can be calculated analytically\cite{extremalds} to be:
\begin{equation}\label{eq:extremalSdS}
\text{extremal: } \omega_n / \kappa_c = - (n-1/2) i + \sqrt{l(l+1) - 1/4} \, , \quad n = 1, 2, ... \, , 
\end{equation}

The other limit, $M/L \rightarrow 0$, has two qualitatively different interpretations: it can be reached by taking $M \rightarrow 0 $ at fixed $L$, corresponding to the limit of empty de Sitter, 
or $L \rightarrow \infty$ at fixed $M$, corresponding to the limit of the Schwarzschild black hole in asymptotically flat spacetime.
The former has analytic quasinormal modes\cite{puredS}:
\begin{equation}\label{eq:dSanalytic}
\text{de Sitter: } \omega_1 / \kappa_c = - l i ; \quad \omega_n = -(l + n) i\, ,  \quad n =  2, 3, ... \, .
\end{equation}
For $l = 0$ this gives a zero mode, which is present for any $M/L$, reflecting the fact that $\phi$ can be shifted by a constant.

There is no analytic result for the quasinormal modes of Schwarzschild black hole in asymptotically flat spacetime, but of course they have been computed numerically in e.g. \cite{flatS} and in the previous section.

For small $M/L$, we expect the equilibration process of the full space time not to be influenced by the presence of a small black hole, 
and conversely we do not expect the equilibration of the black hole to be significantly different from the same black hole in asymptotically flat spacetime.
So we expect to see two decoupled sets of quasinormal modes which are small deformations of those of empty de Sitter and of Schwarzschild in asymptotically flat spacetime.

In Fig. (\ref{fig:SdSim}) we show the imaginary part of the quasinormal modes, for $l=0$ and for the full range of $M/L$.
In both cases, blue lines correspond to complex frequencies, while red lines correspond to purely imaginary frequencies.
On the left plot, the modes are normalized with respect to $\kappa_c$, and at $M/L \approx 0$ we indeed see the pure de sitter modes of Eq. (\ref{eq:dSanalytic}). 
As $M/L$ grows they move down the complex plane, but they stay on the imaginary axis for the whole range of $M/L$.
The extremal values of Eq. (\ref{eq:extremalSdS}) are approached by the other set of modes, which are complex in general but become purely imaginary in the extremal limit $M/L \rightarrow 1/(3\sqrt{3})$.

In Fig. (\ref{fig:SdSim}b) the same data is presented, but now normalized as $\omega M$.
In these units, the complex modes approach those of the Schwarzschild black hole in flat space, as can be seen by comparing with Table (\ref{fig:flatmodes}b), and as indicated by the blue dashed lines.
All of the imaginary modes now disappear in the limit, simply because $M$ goes to zero.
In this limit it becomes numerically difficult to compute the complex modes, as there are so many smaller purely imaginary modes.

\begin{figure}[t]
  \centering
  \includegraphics[width=\textwidth]{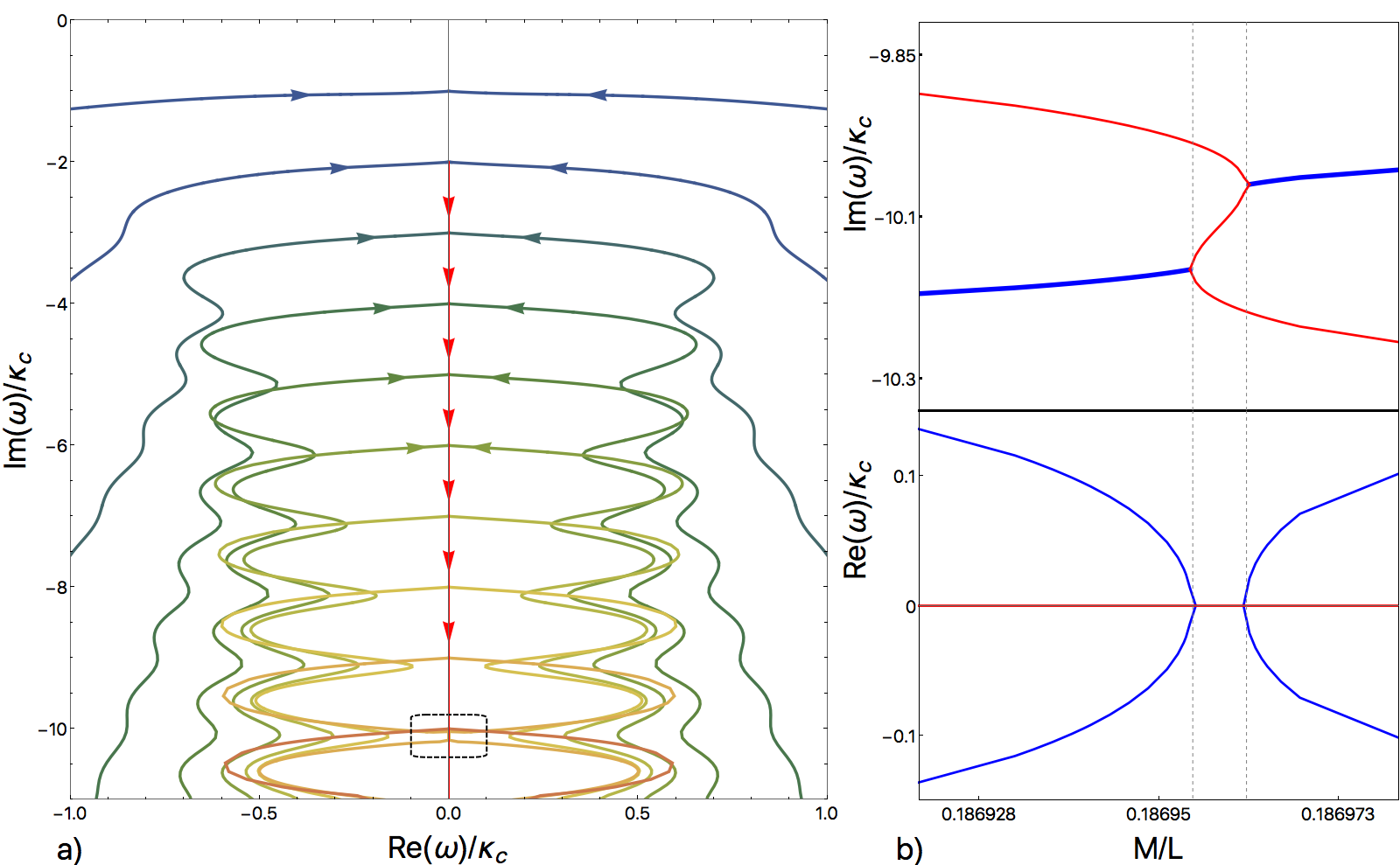}
  \caption{
  Massless scalar quasinormal modes of the asymptotically de Sitter Schwarzschild black hole in 4 dimensions for $l=0$.
  a) Shown for all $M/L$ simultaneously, arrows indicate increasing $M/L$.
  The red line indicates the purely imaginary modes, moving down the imaginary axis. 
  All other lines are the complex modes, with a different color for each mode number, which move up the complex plane.
  b) Zoomed in on the region indicated in a), now as a function of $M/L$ with real and imaginary part shown separately.
  }
  \label{fig:jellyfish}
\end{figure}

Note that in this asymptotically flat limit, the boundary condition at the cosmological horizon approaches the boundary condition of outgoing waves in asymptotically flat spacetime continuously.
This is most easily seen in different coordinates, $ds^2 = - f(r) dt^2 + f(r)^{-1} dr^2 + r^2 \left( d\theta^2 + \sin^2 (\theta) d\phi^2 \right)$, where $f$ is the same as in Eq. (\ref{eq:fSdS}), with $r = 1/u$.
Defining the ``tortoise'' coordinate $r_\star$ by $dr_\star/dr = 1/f(r)$ and $\phi(t,r,\theta,\phi) = e^{-i \omega t} r^{-1} Y_{l m}(\theta,\phi) \psi(r)$ we obtain the Schr\"{o}dinger form of the quasinormal mode equation,
\begin{equation}\begin{split}\label{eq:schr}
0 &= \partial_{r_\star}^2 \psi + \left( \omega^2 - V(r) \right) \psi \, , \\
V(r) &=  f(r) \left[ \frac{l(l+1)}{r^2} + \frac{f^\prime(r)}{r} \right]  \, .
\end{split}\end{equation}
The Schr\"{o}dinger potential $V(r)$ approaches zero both at the cosmological horizon of Schwarzschild de Sitter and as $r \rightarrow \infty$ in the asymptotically flat case, so that expanding $\psi$ near these points we find in both cases $\psi(r) = \exp\left(\pm i \omega r_\star \right)$, where we have to choose the plus sign to obtain the outgoing wave.

In the anti-de Sitter case, the Schr\"{o}dinger potential no longer approaches zero at the boundary, 
and so in the flat limit the boundary condition does not approach that of the asymptotically flat Schwarzschild black hole.
This is why in this case, discussed in section \ref{sec:AdS}, there isn't a set of modes converging to the asymptotically flat Schwarzschild values.

So the limits match, but what happens in between is also interesting.
Firstly, for small $M/L$ the purely imaginary mode coming from de Sitter is dominant, so the late time behavior of a small perturbation (that is sufficiently spread out to excite both sets of modes) is purely damped at late times.
At $M/L \approx 0.051$ there is a crossover, where the complex mode becomes dominant and late time behavior is a damped oscillation.
This crossover occurs at $M/L \approx 0.090, 0.047, 0.032 $ for $l = 1, 2, 3$, so for $l=1$ the imaginary mode dominates in the largest range, which matches the fact that the dominant mode in pure de Sitter is that of $l=1$ (excluding the nondynamical zero mode at $l=0$).

For larger $M/L$ we no longer expect two decoupled sets of modes, and in fact we observe an interaction between the two sets, which produces the oscillations seen in Fig. (\ref{fig:SdSim}).
This can be seen more clearly in Fig. (\ref{fig:jellyfish}a), where we show the spectrum in the complex plane, plotted for all $M/L$ at the same time, with arrows indicating increasing $M/L$.
What happens is that as a purely imaginary mode moves down the axis and a complex mode reaches a similar imaginary part, the complex mode moves towards the axis, as if attracted by the imaginary mode, 
and moves away again once the difference in imaginary parts increases again.

The ``attraction'' becomes stronger for higher complex modes, and for each it is strongest when it approaches the lowest imaginary mode.
This makes intuitive sense, as the larger the mode number, the higher the value of $M/L$ when it crosses the lowest imaginary mode.

The consequence of this interaction is that black holes of certain masses, where the imaginary parts of a complex mode and a purely imaginary mode are equal, will oscillate slightly less.
It would be hard to detect this in a real evolution though, as the effect is quite small for the lower modes, and it occurs at a different mass for each mode.

Starting from the 9th lowest complex mode the complex modes actually reach the imaginary axis and hit the imaginary mode before moving away from it again, in the region indicated by the box in Fig. (\ref{fig:jellyfish}a).
What happens is more clearly seen in Fig. (\ref{fig:jellyfish}b), which shows only the 9th mode in the region of this box, plotting the real and imaginary part separately as a function of $M/L$.
Since $\phi$ is real, complex modes must occur in conjugate pairs, but once they become purely imaginary they can split.
The conjugate pair moves towards the real axis and splits into two purely imaginary modes when they hit the axis.
One of these imaginary modes moves up the axis towards the original purely imaginary mode, which it then pairs up with to become complex again.
The other continues down the axis and stays imaginary.

For higher $l$ these interactions are weaker, with $l=1$ still showing some oscillations but from $l=2$ on they appear to be absent.
Apart from that the picture at larger $l$ is qualitatively similar to what has been discussed.
In Table (\ref{tab:comparison}) we present some numerical values. 
These match the values of \cite{dsnumeric}, which used a sixth order WKB approximation, except that we find the additional infinite set of purely imaginary modes, which have not been reported before in the literature.
The WKB approximation is known to miss overdamped modes\cite{reviewBertiCardosoStarinets}, which in some cases misses the mode that is physically the most relevant. 

These modes do fit nicely with \cite{evolutionSdS}, which did an explicit time evolution of a scalar perturbation of a black hole in de Sitter with $\Lambda \ll 1$ (i.e. very small $M/L$).
They observe an initial exponential decay with approximately the asymptotically flat Schwarzschild black hole QNM, and a late time exponential decay with approximately the empty de Sitter QNM.
Here we showed how these QNMs are modified for larger $M/L$, and that this picture only holds up to some crossover value where the complex mode is dominant.

Although we did not compute electromagnetic or metric perturbations it seems clear that there also one expects purely imaginary modes as continuations of pure de Sitter modes, just as for the scalar.
One might then also expect similar oscillations in the complex modes at the lowest $l = s$. 
It is therefore interesting to look at \cite{Konoplya:2004uk} in this light, which studies the high overtones of these electromagnetic and metric perturbations.
It is shown that at fixed $\Lambda M^2 = 0.02$ and for large overtone $n$, $Re(\omega_n)$ oscillates as a function of $Im(\omega_n)$. 
This might be similar to the oscillation seen in Fig. (\ref{fig:jellyfish}a), since taking a snapshot at constant $\Lambda M^2$ would still produce oscilaltory behavior as a function of $n$,
although it is difficult to give a more precise relation, since we consider only scalar modes, at much lower $n$.

\section{Einstein-Maxwell-scalar model}\label{sec:EMD}
To illustrate the method in a more complex example, we will look at the quasinormal modes of the $4+1$-dimensional Reissner-Nordstr\"{o}m anti-de Sitter black brane.
Before we specialize to this case we will first derive the equations in a much more general setting, namely any Einstein-Maxwell-scalar black brane that is homogeneous and isotropic (and ofcourse time-independent), in any number of dimensions.

We consider the following action,
\begin{equation}\label{eq:SEMD}
S = \frac{1}{2\kappa} \int d^{d+1}x\sqrt{-g} \left( R -\Lambda -  \xi (\partial \phi)^2 -\frac{1}{4} Z(\phi) F^2 - V(\phi)  \right) \, ,
\end{equation}
where $F = d A$ is the field strength of a $U(1)$ gauge field $A$, which has its kinetic term modified by a function $Z(\phi)$ of a scalar field $\phi$, we have a cosmological constant $\Lambda$ and a potential $V(\phi)$ for the scalar field, 
and finally $\xi$ is an arbitrary normalization factor for the scalar field kinetic term.
We work in units where $\kappa = 1/(8\pi G) = 1/2$.

The equations of motion derived from this action are as follows,
\begin{equation}\begin{split}
0 &= R_{\mu\nu} - \frac{1}{2} g_{\mu\nu} R + \frac{1}{2} g_{\mu\nu} \Lambda + \frac{1}{2} \xi g_{\mu\nu} \left(\partial\phi\right)^2 - \xi \partial_\mu \phi \partial_\nu \phi + \frac{1}{2} g_{\mu\nu} V(\phi) \, \\
&+ \frac{1}{8} g_{\mu\nu} Z(\phi) F^2 - \frac{1}{2} Z(\phi) F_{\mu\rho} F_\nu{}^\rho \, , \\
0 &= \nabla_{\mu} \left( Z(\phi) F^{\mu\nu} \right) \, , \\
0 &= \frac{2\xi}{\sqrt{-g}} \partial_\mu \left(\sqrt{-g} \partial^\mu \phi \right) - \frac{1}{4} Z^\prime(\phi) F^2 - V^\prime(\phi) \, .
\end{split}\end{equation}

We specialize to a homogeneous, isotropic and time-independent planar black hole background, allowing us to use the following ansatz,
\begin{equation}\begin{split}\label{eq:EMDansatz}
ds^2 &= - f(u) dt^2 + 2 \zeta(u) dt du + S(u)^2 dx^2_{(d-1)} \, , \\
A &= A_t(u) dt \, , \\
\phi &= \phi(u) \, .
\end{split}\end{equation}

In these coordinates the temperature of any black hole with a horizon at $u_h$ is given by $T = 1/(2\pi) \sqrt{-1/2 \left(\nabla_\mu \chi_\nu\right) \left( \nabla^\mu \chi^\nu \right) }|_{u=u_h} = f^\prime(u_h) / (4 \pi \zeta(u_h) )$.

In order to bring the quasinormal mode equations into a useful form, we follow the approach of \cite{KovtunStarinets}, which we generalize from pure gravity to this rather general Einstein-Maxwell-scalar case.

We start by fluctuating all of the fields as plane waves, with momentum $q$ in the $x$ direction,
\begin{equation}\begin{split}\label{eq:EMDfluct}
\delta g_{\mu\nu} &=  e^{-i\left( \omega t - q x\right)} h_{\mu\nu}(u) \, , \\
\delta A_\mu &=  e^{-i\left( \omega t - q x\right)} a_\mu(u) \, , \\
\delta\phi &=  e^{-i\left( \omega t - q x\right)} \delta\phi(u) \, .
\end{split}\end{equation}

In what follows we will denote spatial indices other than $x$, so transverse to the momentum, with $\alpha \neq \beta$.

We can classify all of these fluctuations according to their helicity, or their transformation properties under the remaining $SO(2)$ symmetry of rotations around the $x$-asix.
\footnote{For $d>4$ there is a larger $SO(d-2)$ symmetry, but it suffices to look at the same $SO(2)$ and take the same representatives in all the channels.}
This immediately groups the fluctuation equations into three mutually decoupled groups, also called channels, with helicities 0, $\pm 1$ and $\pm 2$ and containing the fields,
\begin{equation}\begin{split}
\text{(helicity $\pm$ 2):}&\quad  h_{\alpha,\beta}, \, h_{\alpha,\alpha} - h_{\beta,\beta} \, , \\
\text{(helicity $\pm$ 1):}&\quad  a_\alpha, \, h_{t,\alpha}, \, h_{r,\alpha}, \, h_{x,\alpha} \, , \\
\text{(helicity 0):}&\quad \delta\phi, \, a_t, \, a_r, \, a_x, \, h_{t,t}, \, h_{t,r}, \, h_{t,x}, \, h_{r,r}, h_{r,x}, \, h_{x,x}, \, h   \, ,
\end{split}\end{equation}
where for helicity $0$ we defined $h \equiv \Sigma_{\alpha=4}^{d+1} h_{\alpha\alpha}/(d-2) $.

Note that for $d=3$ there is only one transverse spatial direction, so there are no helicity 2 fields. For any higher $d$ the decomposition is as above.

Since the quasinormal modes that we want to calculate are gauge invariant quantities, we expect to be able to find gauge invariant equations for them.
In order to do this, we have to express the fluctuations in gauge invariant combinations, under infinitesimal diffeomorphisms $\xi_\mu(u,t,x) = e^{-i (\omega t - q x)} \xi_\mu(u)$, and infinitesimal $U(1)$ gauge transformations $\lambda(u,t,x) = e^{-i(\omega t - q x)} \lambda(u)$, which act on the fluctuations as
\begin{equation}\begin{split}\label{eq:diffeo}
\delta h_{\mu\nu} &= -\nabla_\mu \xi_\nu - \nabla_\nu \xi_\mu \, , \\
\delta a_\mu &=   - \partial_\mu \lambda - \xi^\rho \nabla_\rho A_\mu - A_\rho \nabla_\mu \xi^\rho \, , \\
\delta \phi &= -\xi^\rho \nabla_\rho \phi \, ,
\end{split}\end{equation}
where all covariant derivatives are taken only with respect to the background metric (other contributions would be second order).

By simply taking an arbitrary linear combination of all the fluctuations in a given channel, performing the above gauge transformation on this combination and demanding the variation to vanish we find the following gauge invariant variables,
\begin{equation}\begin{split}\label{eq:gaugeindepvars}
\text{(helicity $\pm$ 2):}& \quad Z_3 = h_{\alpha,\beta} \, , \\
\text{(helicity $\pm$ 1):}& \quad Z_1 = q h_{t,\alpha} + \omega h_{x,\alpha} \, , \\
& \quad  E_\alpha = a_\alpha \, , \\
\text{(helicity 0):}& \quad Z_2 = q^2 h_{t,t} + 2 \omega q h_{t,x} + \omega^2 h_{x,x} +  \left(- \omega^2 +  q^2 \frac{f^\prime}{2 S S^\prime } \right) h \, , \\
& \quad E_x = q a_t + \omega a_x - q \frac{A_t^\prime}{2 S S^\prime} h  \, , \\
& \quad  Z_\phi = \delta\phi - \frac{\phi^\prime}{2 S S^\prime} h\, .
\end{split}\end{equation}
In the helicity 2 case $h_{\alpha,\alpha} - h_{\beta,\beta}$ is also gauge invariant and yields the same equation.

The combinations of Eq. (\ref{eq:gaugeindepvars}) are gauge invariant for any dimension $d \geq 4$.
For $d=3$ the only difference is that there is no helicity 2 component as noted above, but the other combinations remain identical.
These are the unique (up to taking linear combinations) gauge invariant combinations of the fluctuations, though if one allows for their radial derivatives as well there are more possibilities \cite{masterequations}.

Substituting the gauge invariant fluctuations into the fluctuation equations, they can be completely decoupled from the gauge-dependent ones, 
again simply by taking an arbitrary linear combination of the equations in a given channel and demanding the coefficients of the gauge-dependent functions and their derivatives to vanish.
We stress that it is not necessary to make any gauge choice in order to do this, although it simplifies the process.

In the general case the gauge invariant variables in the same channel do remain coupled however, resulting in a single decoupled equation for the helicity 2 channel, two coupled equations for helicity 1, and three coupled equations for helicity 0.

Very schematically, not writing any of the coefficients, the equations take the following form, for the helicity 2 channel,
\begin{equation}
0 = Z_3(u) + Z_3^\prime(u) + Z_3^{\prime\prime}(u)\, , 
\end{equation}
for the helicity 1 channel,
\begin{equation}\begin{split}
0 &=  E_y(u) + E_y^\prime(u) + Z_1(u) + Z_1^\prime(u) + Z_1^{\prime\prime}(u) \, , \\
0 &= Z_1(u) + Z_1^\prime(u) + E_y(u) + E_y^\prime(u) + E_y^{\prime\prime}(u)  \, , 
\end{split}\end{equation}
and for the helicity 0 channel,
\begin{equation}\begin{split}
0 &= Z_\phi(u) + E_x(u) + E_x^\prime(u) + Z_2(u) + Z_2^\prime(u) + Z_2^{\prime\prime}(u)\, , \\
0 &= Z_2(u)  + Z_2^\prime(u) + E_x(u)  + E_x^\prime(u)   +  Z_\phi(u) + Z_\phi^\prime(u) + Z_\phi^{\prime\prime}(u)\, , \\
0 &= Z_2(u) + Z_2^\prime(u)  + Z_\phi(u) + Z_\phi^\prime(u) +   E_x(u) + E_x^\prime(u) + E_x^{\prime\prime}(u)\, .
\end{split}\end{equation}
Each term has a coefficient that in general depends on $u$, the frequency $\omega$ and the momentum $q$, and any parameters of the background solution.

Note that even though the original coupled equations are only quadratic in $\omega$, through the decoupling higher powers of $\omega$ arise.
The equations go up to $\omega, \omega^3$ and $\omega^5$ for helicity 2, 1 and 0 respectively, though the imposing of boundary conditions may raise this further, as we saw in sections \ref{sec:flat} and \ref{sec:SdS}.

At zero momentum, most equations decouple. The equations for $E_\gamma (\gamma=x, y, z)$ become identical to eachother, as do those for $Z_\gamma (\gamma=1, 2, 3)$, however the fluctuation $Z_2$ in general still occurs in the equation for $Z_\phi$.
Furthermore, if the background gauge field vanishes the equations for $E_\gamma (\gamma=x, y, z)$ decouple, since the action is quadratic in the gauge field,
and if the background value of the scalar vanishes, the equation for $Z_\phi$ decouples only if the potentials $V(\phi), Z(\phi)$ are at least quadratic in $\phi$.

\subsection{Reissner-Nordstr\"{o}m anti-de Sitter black brane}
We now apply the previous to the asymptotically AdS$_5$ Reissner-Nordstr\"{o}m black brane.
We will encounter most of the numerical difficulties of the general Einstein-Maxwell scalar black brane, 
while still having an analytic background and even some analytic control on the quasinormal modes.
This allows us to demonstrate the method in a more complicated case while also showing that it gives correct results.

The background, in terms of the ansatz of Eq. (\ref{eq:EMDansatz}), is as follows,
\begin{equation}\label{eq:RNbackground}\begin{split}
f(u) &= u^{-2} \left( 1 - M u^4 + Q^2 u^6 \right)\, , \\
S(u) &=  u^{-1}\, , \\
\zeta(u) &= - u^{-2} \, , \\
A_t(u) &=  \sqrt{3} Q \left( u^2 - 1 \right)\, , \\
\end{split}\end{equation}
and ofcourse $\phi = 0$.
We fix the horizon to be at $u=1$, which fixes $M = 1 + Q^2$. The temperature of this brane is $T = (1 - Q^2/2)/\pi$, so the extremal solution is $Q_{\text{extr}} = \sqrt{2}$, which has a vanishing temperature.
It has an entropy density $s = 1/(4\pi)$ and a chemical potential $\mu = \sqrt{3} Q$.
This background has a single dimensionless ratio, which we choose to take as $\tilde{Q} \equiv Q/Q_{\text{extr}}$.

We simply plug this background into the generally derived quasinormal mode equations. As before, the boundary condition of ingoing modes at the horizon is enforced automatically by using Eddington-Finkelstein coordinates.
The only thing we still need to do is fix the boundary conditions at the boundary $u = 0$, where we must set the non-normalizable mode to zero.
For each gauge-independent fluctuation, the solutions near the boundary, and the rescaling we do to enforce this boundary condition, are
\begin{equation}\label{eq:rescalingRN}\begin{split}
Z_\gamma(u) &\propto s u^{-2} (1 + ...)  + v u^2 (1 + ...) \, ; \quad\quad\quad Z_\gamma(u) = u \tilde{Z}_\gamma(u) \quad (\gamma = 1, 2, 3, \phi) \, ,\\
E_\gamma(u) &\propto s u(1 + ...) + v u^2 (1 + ...) \, ; \, \, \, \, \, \,  \quad\quad\quad E_\gamma(u) = u \tilde{E}_\gamma(u) \quad (\gamma = x, y) \, ,\\
\end{split}\end{equation}
where each fluctuation is now rescaled such that the non-normalizable mode diverges, while the normalizable mode approaches 0 linearly.

\begin{figure}[t]
  \centering
  \includegraphics[width=\textwidth]{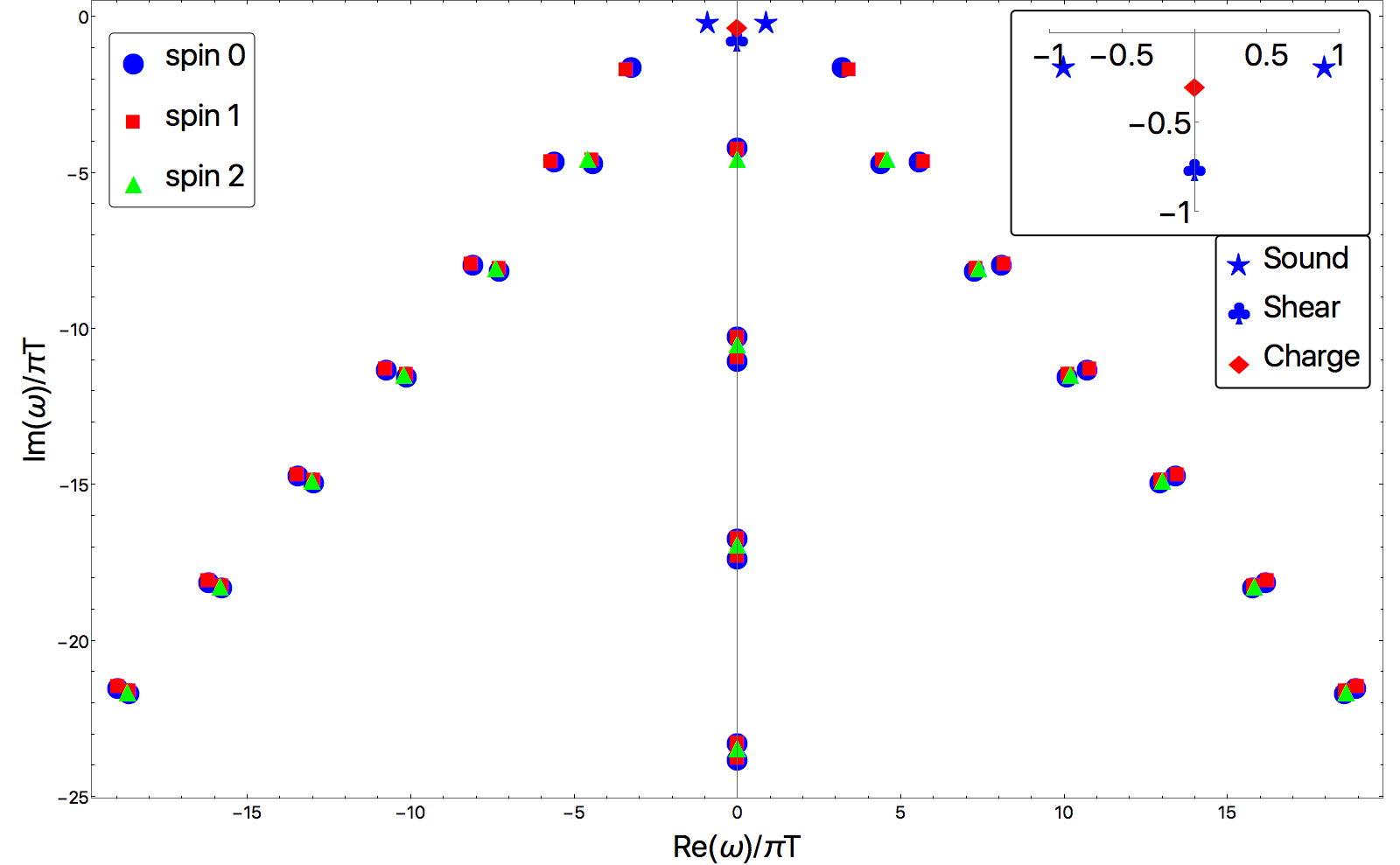}
  \caption{Quasinormal mode spectrum of the Reissner-Nordstr\"{o}m anti-de Sitter black brane, with charge $Q/Q_{\text{extr}} = 1/2$ and momentum $q/\pi T = 1.5$.}
  \label{fig:highermodesRN}
\end{figure}

Having made these redefinitions we are ready to compute the quasinormal modes.
In Fig. (\ref{fig:highermodesRN}) we show the full quasinormal mode spectum of all the channels, for the background charge $Q/Q_{\text{extr}} = 1/2$ and momentum of the perturbations $q/\pi T = 1.5$.
Modes of the different channels nearly overlap. This can be understood from the fact that, as we mentioned in the previous section, at zero momentum the equations for all the $E_\gamma$ become identical to eachother, 
giving the nearly overlapping spin 0 and spin 1 modes coming from $E_x$ and $E_y$. The equations for all the $Z_\gamma$ also become identical, giving the nearly overlapping spin 0, 1 and 2 modes.
It is somewhat remarkable though that the modes are still so close at the reasonably high $q/\pi T = 1.5$.

In Table (\ref{tab:highermodes}) we give the numerical values of the lowest 10 quasinormal modes in each channel, for $Q/Q_{\text{extr}} = 1/2$ and $q/\pi T = 1.5$ as in Fig. (\ref{fig:highermodesRN}).

\begin{figure}[t]
  \centering
  \includegraphics[width=\textwidth]{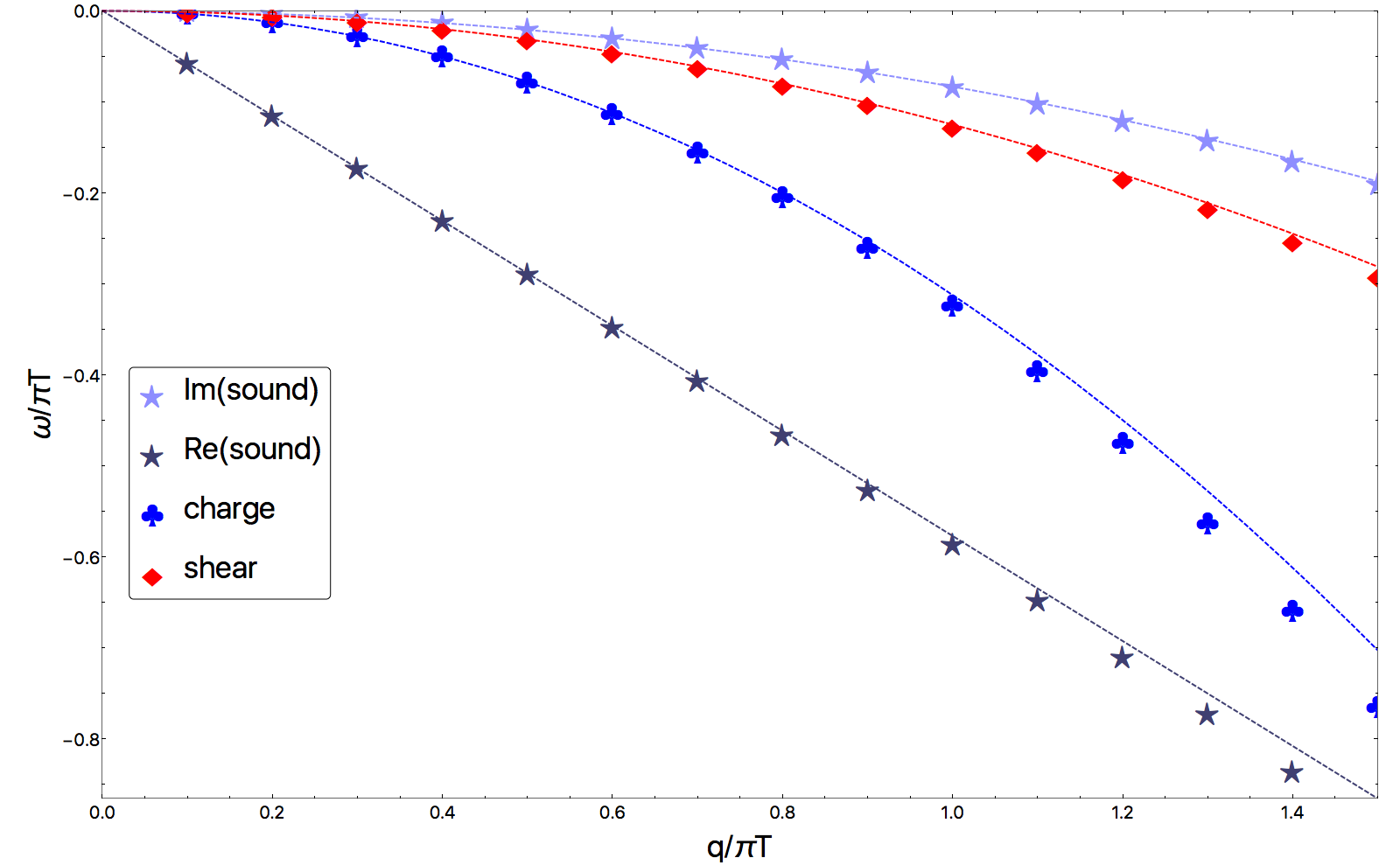}
  \caption{
  Hydrodynamic modes of the Reissner-Nordstr\"{o}m anti-de Sitter black brane, with charge $Q/Q_{\text{extr}} = 1/2$.
  Points are numerically computed quasinormal modes, dashed lines are the dispersion relations of Eq. (\ref{eq:dispersion2}).}
  \label{fig:hydromodes}
\end{figure}

What depends much more strongly on the momentum are the three modes near $\omega = 0$, shown enlarged in the inset.
These are the hydrodynamic modes, which have the property that $\omega(q\rightarrow 0)\rightarrow 0$.
The spin 1 channel contains the shear mode, while the spin 0 channel contains the sound mode, and a mode governing charge diffusion.

In Fig. (\ref{fig:hydromodes}) we show the momentum-dependence of these hydrodynamic modes, again for $Q/Q_{\text{extr}} = 1/2$.
At low momentum these modes are given by the dispersion relations(see e.g. \cite{hydro1}), as: 
\begin{equation}\label{eq:dispersion}\begin{split}
\text{shear:} \quad\quad&\omega(q) = - i \frac{\eta}{\epsilon + P} q^2 \, , \\
\text{sound:} \quad\quad&\omega(q) =  \pm v_s q - \frac{i}{2} \frac{1}{\epsilon + P}\left(\zeta + \frac{4}{3} \eta \right) q^2\, , \\
\text{charge:} \quad\quad&\omega(q) = - \frac{i}{2} \frac{D}{\epsilon + P} q^2\, , 
\end{split}\end{equation}
where $\epsilon$ and $P$ are the energy and pressure, $\eta$ and $\zeta$ the shear and bulk viscosity, $v_s^2 = \partial P / \partial \epsilon$ is the speed of sound and $D$ is the charge diffusion, normalized to be dimensionless and 1 at $Q=0$.

The shear viscosity is universally fixed, for any two derivative gravitational action, to be $\eta/s = 1/(4\pi)$\cite{etaovers}.
Furthermore the theory we are looking at is conformal, so that the bulk viscosity vanishes and $v_s^2 = 1/(d-1) = 1/3$.
This means that the sound mode is also completely fixed. 
While the charge diffusion constant $D$ is not fixed by $\eta/s$ and conformality, it can be calculated analytically \cite{chargedbranes,chargediffusion}, so that the three hydrodynamic modes become,
\begin{equation}\label{eq:dispersion2}\begin{split}
\text{shear:} \quad\quad&\frac{\omega(q)}{\pi T} = - \frac{i}{4} \frac{1-\tilde{Q}^2}{1+2\tilde{Q}^2} \left(\frac{q}{\pi T}\right)^2 \, , \\
\text{sound:} \quad\quad&\frac{\omega(q)}{\pi T} = \pm \frac{1}{\sqrt{3}}\frac{q}{\pi T} - \frac{i}{6} \frac{1-\tilde{Q}^2}{1+2\tilde{Q}^2} \left(\frac{q}{\pi T}\right)^2 \, , \\
\text{charge:} \quad\quad&\frac{\omega(q)}{\pi T} = - \frac{i}{2} \frac{1-\tilde{Q}^2}{1+2\tilde{Q}^2} \left(1 + \tilde{Q}^2 \right) \left(\frac{q}{\pi T}\right)^2 \, 
\end{split}\end{equation}
where we used $\epsilon + P = 4M = 4 (1 + 2 \tilde{Q}^2)$, following from a standard renormalization procedure\cite{renorm}.

In Fig. (\ref{fig:hydromodes}) Eqs. (\ref{eq:dispersion2}) are plotted as dashed lines going through the numerically computed points.
In each case the dispersion relations describe the modes well at low momenta, although we can already see the higher order corrections around $q/\pi T \approx 1$.
We also check that the dispersion relations of Eq. (\ref{eq:dispersion2}) are followed for any other $Q$.

\section{Discussion} 
The example of the Schwarzschild-de Sitter black hole illustrates one of the main advantages of the method we employed.
As one of the simplest black holes, its quasinormal modes have been computed already in 1990\cite{firstSdS}, 
yet all this time the infinite set of purely imaginary modes, which depending on the black hole mass may even be the dominant mode, have been missed.
This is either because the approximation used misses this type of modes (e.g. the WKB approximation\cite{WKB}), or in other cases. e.g. with Leaver's method\cite{leaver}, because the method requires an initial guess, effectively giving only what's expected.
We checked that even the thirteenth order WKB approximation with Pad\'e resummation of \cite{WKBhighorder}, while giving very accurate results for the complex modes, still completely misses the purely imaginary modes.

The method used here is completely a-priori, it has no guesses, no assumptions other than that the eigenfunctions be analytical, and no approximations other than the grid-size, which can easily be varied.

Two additional examples, rederiving the QNM's of \cite{KovtunStarinets} and \cite{lifshitzevolution}, are included in \cite{package}.
Although we did not discuss a case where the background is known only numerically, it should be clear that as long as the numerical background is known to a high enough precision this will not give any problems, 
as an analytical background has to be converted to a numerical one in any case.
In \cite{scalarinstability} we did use the package successfully for a numerical background, and the same method as we use here has been used with numerical backgrounds for instance in \cite{numericalbackground1,numericalbackground2,chargedbranes},
and in analytic but more involved situations for instance in \cite{Hartnett:2013fba,Dias:2011jg,Dias:2010gk,Dias:2010ma,Dias:2010maa}.

In an upcoming work\cite{RNdSSCC} we will study the QNM's of the Reissner-Nordstr\"{o}m black hole in asymptotically de Sitter spacetime using \cite{package}, in particular in relation to strong cosmic censorship.

There are several possible extensions to the current method.
A limitation of the current approach is that the equation, after the rescaling to implicitly solve the boundary conditions, has to have a polynomial dependence on the frequency. 
This is not always the case, see for instance \cite{QNMnonpoly}. There are two main ways around this issue.
The most closely related one is to take the determinant of the matrix representing the full QNM equation, as a function of the frequency, and look for zeroes by evaluating it on numerical points in the complex plane \cite{QNMbydet}.
Another way, developed and used in \cite{QNMbyNR}, is to use the Newton-Raphson method. 
This can be advantageous even when not strictly necessary, as it allows one to efficiently track a single mode as a parameter is varied \cite{Santos:2015iua, Dias:2015wqa, Dias:2014eua}.

Another extension that is possible and would be interesting is to nonhomogeneous backgrounds.
The main complication there is that the matrices will get a lot larger, by a factor of the size of the grid in the extra direction.
All these extensions would fit nicely in the current framework and are worth adding in the future.

We hope that the present package will be of use to the community. 
We have certainly found it very convenient in our own work, and believe others could benefit from it.
In addition, we encourage anyone to contribute with optimizations or bug fixes or extra features, it is hosted on GitHub, which facilitates such collaboration.

\section{Acknowledgements}
We want to thank especially L. Yaffe and the organizers of the 2014 Mathematica summer school on theoretical physics in Porto, as this package was inspired by an exercise by L. Yaffe at this school \cite{qnmYaffe}. 
We also want to thank W. van der Schee and W. Sybesma for using earlier versions of this package and helping fix the occasional bugs and other improvements, and U. Gursoy, J. M. Magan and S. Vandoren for collaborations on different projects during which this package was used and developed. 
We thank N. Kaplis, M. Ammon and O. Dias for helpful discussions regarding the numerics, J. Matyjasek for help with his method, V. Cardoso  for interesting discussions on the quasinormal modes of the Schwarzschild and Schwarzschild-de Sitter black holes and Jason Harris for comments and suggestions on the code.

This work was supported by the Netherlands Organisation for Scientific Research (NWO) under VIDI grant 680-47-518, and the Delta-Institute for Theoretical Physics (D-ITP) that is funded by the Dutch Ministry of Education, Culture and Science (OCW).

\appendix

\section{Benchmarks}\label{app:benchmarks}
Here we provide some benchmarks on the performance, in terms of speed and accuracy.

We will consider again the massless scalar in Schwarzschild-AdS, Eq. (\ref{eq:qnmfinite}). 
When it comes to timing we expect this equation to be fully representative, the speed should at least to a good approximation be independent of the details of the equation. 
One just has to keep in mind that for $N_{\text{eq}}$ coupled equations with the maximal power of the frequency being $p$ the matrix size scales as $N_{\text{eq}} \, p$.

\begin{figure}[htb]
  \centering
  \begin{minipage}[c]{0.5\textwidth}
    \centering
    \includegraphics[width=\textwidth]{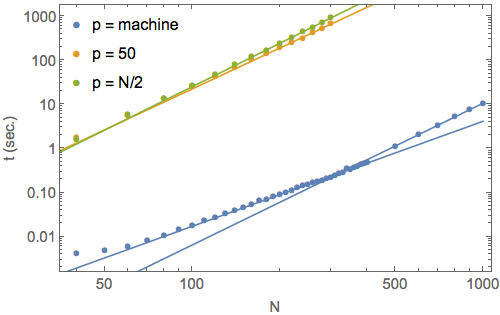}
  \end{minipage}
  \begin{minipage}[c]{0.4\textwidth}
    \centering
\begin{tabular}{c|c|}
$(N,p)$        &  t (seconds)             \\ \hline\hline
(40,0)           &  0.004       \\ \hline
(100,0)         &   0.018      \\ \hline
(250,0)         &   0.16        \\ \hline
(1000,0)       &  8.9           \\ \hline\hline
(40,20)         &  1.6           \\ \hline
(100,50)       &   26           \\ \hline
(250,125)     &   568         \\ \hline
(1000,500)   &  14 hours  \\ \hline
\end{tabular}
  \end{minipage}
\caption{
Benchmark for performance and its dependence on the grid size $N$, for various fixed values of the precision $p$. In addition the ... line has precision $p = N/2$.
Lines are power law fits of $t(N) = t_0 N^q$, finding $q \approx 2.377$ using machine precision, $q \approx 3$ using constant high precision and $q \approx ... $ using precision $p = N/2$.
On the right a table with several absolute values, running on a single 1,7 GHz Intel i7 Core, with 8GB of RAM.}
\label{fig:speed}
\end{figure}

In Fig. (\ref{fig:speed}) we show a log-log plot of the time $t$ taken to compute the quasinormal modes (in seconds) as a function of the grid size $N$, using machine precision (blue points), a precision of 50 digits (yellow points) and a precision depending on the grid size as $p = N/2$ (green points).
In all cases we find a power law $t(N) = t_0 N^q$. For the computation with machine precision, the power seems to be different for smaller grid sizes ($q \approx 2.38$) and for larger grid sizes ($q \approx 3.2$).
For the case with 50 digits of precision, we find $q \approx 3.1$, while for the one with precision increasing as $N/2$ we find $q \approx 3.3$.
Summarizing, the scaling with $N$ is roughly as $N^3$, except for grids with roughly less than 400 points where it scales as $N^{2.38}$. 
The dependence on precision is very mild, as we can see by comparing the yellow and green points, except in the transition from machine precision to higher precision, which slows the computation down by roughly 2 orders of magnitude.

To the right of Fig. (\ref{fig:speed}) we give a few explicit values of the time taken on a laptop. 
Note that for larger grid sizes and/or using higher precision, the computation of the eigenvalues takes up virtually all of the time, while only for low grid sizes the other steps in the computation, namely the construction of the matrices, become important.
We check that for a grid of 40 points with machine precision, which we consider roughly the smallest grid that can still give valuable results, about $65\%$ of the time is taken up by highly optimized built-in functions of Mathematica (which are programmed in C).
These are the construction of the derivative matrices and the actual computation of the eigenvalues.
This means that in the worst case, where the rest of the computation can be optimized so much as to take a negligible amount of time, the speed gain would be about $35\%$. 
However, since for example the construction of the numerical matrix, which involves computations with vectors (in particular the grid), does take some time, we believe the actual room for improvement to be much smaller.

\begin{figure}[htb]
  \centering
  \begin{minipage}[c]{0.45\textwidth}
    \centering
    \includegraphics[width=\textwidth]{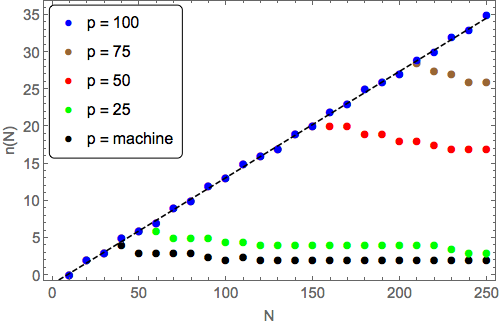}
  \end{minipage}
  \begin{minipage}[c]{0.45\textwidth}
    \centering
    \includegraphics[width=\textwidth]{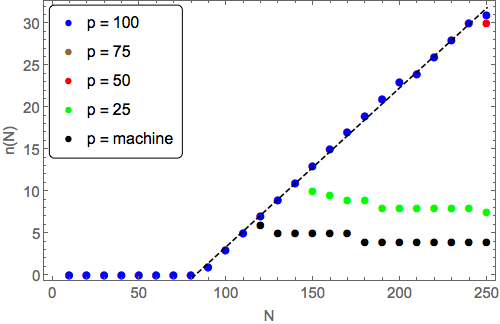}
  \end{minipage}
  \caption{Number of converged modes $n(N)$ found as a function of grid size $N$, for different values of the precision $p$. Left: for the massless scalar at zero momentum, Eq. (\ref{eq:qnmfinite}). Right: for the same scalar but with large momentum $q/\pi T = 160$. Before they branch off all the colors lie on the straight line (below the blue dots). Dashed lines are linear fits to the blue data.}
\label{fig:convergence}
\end{figure}

In Fig. (\ref{fig:convergence}) we show the number of accurately computed modes as a function of the grid size $N$, for different values of precision, as found by comparison with the computation with $N = 300$.
The left figure is for the massless scalar at zero momentum, Eq. (\ref{eq:qnmfinite}).
Note that before the different colors branch off the straight line, they follow it exactly, lying under the blue dots. 
This means that when it comes to the digits of precision used, for a given grid size $N$ it needs to be above some threshold to compute the most accurate modes, but increasing it further has no effect. 
As long as the precision is high enough, the number of accurate modes increases linearly with the grid size, up until the point where the precision is no longer high enough, when the number of modes actually goes down a little with increasing grid size.

In order to see what features we can expect to be generic, independent of the equation we are solving, and which are not, 
we repeat this procedure on the right-hand side of Fig. (\ref{fig:convergence}) for the same scalar but now with a large momentum of $q/\pi T = 160$ (these modes were also computed in \cite{YaffeFuiniUhlemann}, we ofcourse agree on the result). 
This may seem like a trivial change, and on the level of the equation it is, but the solutions become highly oscillatory and thus harder to capture with a few Chebyshev polynomials.

From these two cases, we expect that the two features mentioned above are generic, at least after $N$ is large enough to resolve the first mode.

The details are of course not generic, such as the actual value of $n(N)$ for a specific $N$, or the actual value of the precision $p(N)$ needed, or the coefficient of the linear growth of $n(N)$ (curiously, at high momentum we need less precision than at zero momentum).

\section{Quasinormal mode equations}\label{sec:app2}
In this appendix we write the quasinormal mode equations in their rescaled form which can directly be used in the package.

\paragraph{Quasinormal mode equations}
Here we present the scalar quasinormal mode equations of the Schwarzschild black hole for de Sitter and flat asymptotics, in a form that can directly be used for numerical computation.

For the Schwarzschild black hole, the anti-de Sitter case was given in Eq. (\ref{eq:globalads}) and the asymptotically flat case in Eq. (\ref{eq:flatSchw}).
In de Sitter, using the radial variable $v = (u - u_c)/(u_b - u_c)$ with the cosmological horizon at $v = 0$ and the black hole horizon at $v = 1$ ,
and having rescaled $\phi$ according to Eq. (\ref{eq:dsrescale}) to enforce the correct boundary conditions, the equation is
\begin{equation}\label{eq:SdSfinal}
0 = (1-u_c) C_0 \tilde{\phi}(v) + C_1 \tilde{\phi}^\prime(v) + C_2 \tilde{\phi}^{\prime\prime}(v) \, , 
\end{equation}
with,
\begin{equation}\begin{split}
C_0 &= \bigg( u_c^2 \left(\left(l^2+l+3\right) v-v^3-2\right)+(v-1) u_c^3 \left(-\left(l^2+l+2\right) v+v^2+1\right)+ \\
&v u_c \left(l^2+l-v^2+5 v-6\right)+v^2 \left(l^2+l+v-2\right) \bigg) +\\
&\frac{4 v \omega ^2
   \left(u_c^2+u_c+1\right){}^2 \left(\left(v^2-3 v+2\right) u_c^3-\left(v^2-2\right) u_c^2-\left(v^2-4 v+1\right) u_c+(v-1) v\right)}{\left(u_c-1\right) u_c^2 \left(u_c+2\right){}^2}+ \\
&\frac{2 i \omega  \left(u_c^2+u_c+1\right)
   \left(-2 \left(v^3-2 v+1\right) u_c^2+v \left(-2 v^2+9 v-5\right) u_c+\left(2 v^3-6 v^2+4 v-1\right) u_c^3+v^2 (2 v-3)\right)}{u_c \left(u_c+2\right)} \, , \\
C_1 &= v \left(u_c-1\right) \left(-\left(v^3-3 v+2\right) u_c^2-v \left(v^2-5 v+6\right) u_c+(v-1)^3 u_c^3+(v-2) v^2\right)- \\
&\frac{2 i v \omega  \left(u_c^2+u_c+1\right) \left(\left(-2 v^3+5 v-2\right) u_c^2-2 v \left(v^2-4 v+2\right)
   u_c+\left(2 v^3-6 v^2+5 v-1\right) u_c^3+2 (v-1) v^2\right)}{u_c \left(u_c+2\right)} \, , \\
C_2 &= (v-1) v^2 \left(1-u_c\right) \left(-\left(v^2+v-2\right) u_c^2+(v-1)^2 u_c^3-(v-3) v u_c+v^2\right) \, .
\end{split}\end{equation}

The equations for the general Einstein-Maxwell-scalar backgrounds, and those for the specific case of the Reissner-Nordstr\"{o}m black brane can be found in \cite{package}.

\section{Numerical values}\label{sec:appTables}
In this appendix we provide some quantitative results. All complex modes come with their negative complex conjugate which we do not show.

In Table (\ref{tab:comparison}) we show the lowest lying quasinormal mode of the Schwarzschild-de Sitter black hole discussed in section \ref{sec:SdS}, for $l = 1, 2$, and the second lowest for $l=2$.
We take as parameter $\Lambda M^2 = 3 M^2/L^2$, for ease of comparison with \cite{dsnumeric} which computed the same QNMs with a sixth order WKB approximation.
Results agree to a high precision, except that we find an extra set of purely imaginary modes (in the top left entry this is the dominant mode).
\begin{table}[h!]
\begin{center}
\begin{tabular}{|c||c|c|c|}
 $\Lambda M^2$ & $\omega_1$\text{M (l=1)} & $\omega_1$\text{M (l=2)} & $\omega_2$\text{M (l=2)} \\ \hline\hline
 0.02 & \begin{tabular}{@{}c@{}}0.26028785 - 0.09100254 i \\  - 0.081565496 i\end{tabular} & \begin{tabular}{@{}c@{}}0.43460806 - 0.08857921 i \\  - 0.16326361 i\end{tabular} & \begin{tabular}{@{}c@{}}0.42083979 - 0.26859762 i \\  - 0.32812729 i\end{tabular} \\ \hline
 0.04 & \begin{tabular}{@{}c@{}}0.22468492 - 0.08205129 i \\  - 0.11524810 i\end{tabular} & \begin{tabular}{@{}c@{}}0.38078394 - 0.07875882 i \\  - 0.23084063 i\end{tabular} & \begin{tabular}{@{}c@{}}0.37165302 - 0.23795046 i \\  - 0.46598251 i\end{tabular} \\ \hline
 0.06 & \begin{tabular}{@{}c@{}}0.18536931 - 0.07006012 i \\  - 0.14100253 i\end{tabular} & \begin{tabular}{@{}c@{}}0.32002117 - 0.06684510 i \\  - 0.28266235 i\end{tabular} & \begin{tabular}{@{}c@{}}0.31421964 - 0.20163462 i \\  - 0.57288487 i\end{tabular} \\ \hline
 0.08 & \begin{tabular}{@{}c@{}}0.14040900 - 0.05402319 i \\  - 0.16268011 i\end{tabular} & \begin{tabular}{@{}c@{}}0.24746980 - 0.05190430 i \\  - 0.32632442 i\end{tabular} & \begin{tabular}{@{}c@{}}0.24425463 - 0.15625540 i \\  - 0.66383726 i\end{tabular} \\ \hline
 0.09 & \begin{tabular}{@{}c@{}}0.113996747 - 0.043882692 i \\  - 0.17249210 i\end{tabular} & \begin{tabular}{@{}c@{}}0.20296037 - 0.04255840 i \\  - 0.34608498 i\end{tabular} & \begin{tabular}{@{}c@{}}0.20096528 - 0.12793583 i \\  - 0.70527756 i\end{tabular} \\ \hline
 0.10 & \begin{tabular}{@{}c@{}}0.081589749 - 0.031233160 i \\  - 0.18177480 i\end{tabular} & \begin{tabular}{@{}c@{}}0.14661011 - 0.03068686 i \\  - 0.36477001 i\end{tabular} & \begin{tabular}{@{}c@{}}0.14576370 - 0.09212410 i \\  - 0.74462797 i\end{tabular} \\ \hline
 0.11 & \begin{tabular}{@{}c@{}}0.025490450 - 0.009649425 i \\  - 0.19057630 i\end{tabular} & \begin{tabular}{@{}c@{}}0.046168894 - 0.009631341 i \\  - 0.38253699 i\end{tabular} & \begin{tabular}{@{}c@{}}0.046139455 - 0.028894257 i \\  - 0.78218943 i\end{tabular} \\ \hline
\end{tabular}
\end{center}
\caption{
Scalar quasinormal modes of the 4-dimensional Schwarzschild-de Sitter black hole.
For both the complex modes and the imaginary modes we show the dominant mode for $l = 1, 2$ (first two columns), and the subdominant mode for $l = 2$ (last column).
}
\label{tab:comparison}
\end{table}

In Table (\ref{tab:highermodes}) we show the higher quasinormal modes of Reissner-Nordstr\"{o}m, at $Q/Q_{\text{extr}} = 1/2$ and $q/\pi T = 1.5$, as in Fig. (\ref{fig:highermodesRN}).
The hydrodynamic modes, not shown in the table, are $\omega_{\text{shear}}/\pi T = -0.29171230 i$, $\omega_{\text{sound}}/\pi T = 0.90170632 - 0.19067682 i$ and $\omega_{\text{charge}} / \pi T = -0.76581370 i$.
\begin{table}[h!]
\begin{center}
\begin{tabular}{|c||c|c|c|}
 \text{n} & $\omega_n/\pi T$\text{ (spin 2)} & $\omega_n/\pi T$\text{ (spin 1)} & $\omega_n/\pi T$\text{ (spin 0)} \\ \hline\hline
 1 & 4.5842653 - 4.5115101 i & 3.4208596 - 1.7096808 i & 3.2272521 - 1.5718904 i \\ \hline
 2 &  - 4.5153212 i &  - 4.2748489 i &  - 4.1639636 i \\ \hline
 3 & 7.3858057 - 8.0025743 i & 4.4418736 - 4.6075472 i & 5.5850243 - 4.6094553 i \\ \hline
 4 &  - 10.4611240 i & 5.6998584 - 4.6628700 i & 4.4041390 - 4.6516230 i \\ \hline
 5 & 10.197524 - 11.427669 i & 8.1610285 - 7.9327937 i & 8.0857171 - 7.9030574 i \\ \hline
 6 & 13.013066 - 14.828065 i & 7.2948280 - 8.0706177 i & 7.2694953 - 8.0985328 i \\ \hline
 7 &  - 16.877380 i &  - 10.2859009 i &  - 10.2143993 i \\ \hline
 8 & 15.830644 - 18.217453 i &  - 10.9487578 i &  - 10.9840059 i \\ \hline
 9 & 18.649472 - 21.600858 i & 10.774697 - 11.306532 i & 10.720434 - 11.287166 i \\ \hline
 10 &  - 23.398784 i & 10.129449 - 11.481358 i & 10.110118 - 11.502436 i \\ \hline
\end{tabular}
\end{center}
\caption{Quasinormal modes of the Reissner-Nordstr\"{o}m AdS$_5$ black brane, for $Q/Q_{\text{extr}} = 1/2$ and $q/\pi T = 1.5$ (hydrodynamic modes shown separately).}
\label{tab:highermodes}
\end{table}

\clearpage
\bibliography{bibliography}

\end{document}